# Reassessment of ammonia self- and air-broadened half-widths in the HITRAN database

Ali Elkhazraji[1,2,*]

[1] Department of Mechanical Engineering, King Fahd University of Petroleum & Minerals, Dhahran, Saudi Arabia

[2] Interdisciplinary Research Center of Hydrogen Technologies and Carbon Management, King Fahd University of Petroleum & Minerals, Dhahran, Saudi Arabia

[*]Corresponding Author Email: ali.elkhazraji@kfupm.edu.sa

**Abstract**

Accurate $NH_3$ pressure-broadening parameters are required for reliable line-by-line simulations in a multitude of applications including atmospheric sensing, combustion diagnostics, and planetary/exoplanet applications. Current HITRAN $NH_3$ self- and air-broadened half-widths are largely based on the Nemtchinov rotational correlation and subsequent database rules that clamp or assign default values outside the validated range. This treatment avoids nonphysical extrapolation but suppresses the measured rotational dependence, especially for high-$J''$ transitions. In this work, published laboratory measurements of $NH_3$ self- and air-broadened Lorentz half-widths were compiled and reassessed to develop updated empirical correlations for HITRAN. The collected dataset contains 1317 self-broadened and 1231 air-broadened half-widths at, or reduced to, 296 K, spanning multiple vibrational bands, branches, and rotational quantum numbers. The analysis confirms that the dominant dependence of the linewidths is rotational, rather than branch-, band-, or vibration-inversion-specific. Constrained third-degree polynomial correlations were fitted for $\gamma_{\text{self}}$ and $\gamma_{\text{air}}$ as functions of the branch-dependent rotational index $m$ and $K''$. The regression was weighted by reported measurement uncertainties and constrained to remain positive over the intended HITRAN rotational domain. The new fit reduces the air-broadening mean absolute percentage error (MAPE) from 10.95% for HITRAN2024 to 6.80%. For self broadening, the MAPE decreases from 23.61% for HITRAN2024 to 10.89%. Spectrum-level comparisons with PNNL-database spectra and pure-$NH_3$ measurements show that the updated coefficients improve simulated absorption profiles for both foreign-gas and self-broadened conditions. These results provide a physically constrained, database-ready replacement for the current HITRAN $NH_3$ self- and air-broadening treatment.

## 1. Introduction

Ammonia, $NH_3$, is a key molecule in atmospheric chemistry, combustion science, industrial monitoring, and planetary spectroscopy [1]. In the terrestrial atmosphere, $NH_3$ is the dominant alkaline trace gas [2] and is strongly linked to agricultural activity [3], fertilizer use [4], livestock emissions [5], waste management [6], biomass burning [7], and industrial sources [8]. Because $NH_3$ reacts readily with acidic species to form ammonium aerosols, its abundance affects air quality, nitrogen deposition, ecosystem loading, and atmospheric particulate matter formation [9]. Therefore, $NH_3$ is an important atmospheric trace species which underscores the need for reliable spectroscopic data for satellite and ground-based retrievals [10].

The ability to detect and quantify $NH_3$ has become increasingly important across a multitude of applications. Optical $NH_3$ sensors have been developed for environmental monitoring [11], industrial process control [12], breath analysis [13], waste-water applications [14], and combustion measurements [15,16]. Mid-infrared quantum cascade lasers have enabled sensitive detection [17], while near-infrared diode-laser and cavity-enhanced methods have been used in weaker overtone and combination bands [18]. These applications require accurate line positions, intensities, pressure-broadening coefficients, and line-shape models because calibration-free absorption measurements depend directly on the spectroscopic parameters used in the forward model [19].

$NH_3$ has gained renewed importance as a carbon-free fuel and hydrogen carrier [20]. Recent reviews identify $NH_3$ as a promising alternative fuel because it contains no carbon and can be produced, transported, and stored using existing infrastructure [21]. At the same time, $NH_3$ combustion introduces challenges associated with ignition, flame stability and $NO_x$ formation; such challenges motivate significant ongoing research efforts, highlighting the need for quantitative high-temperature $NH_3$ diagnostics [22]. $NH_3$ is equally important in planetary and exoplanetary spectroscopy. It is a known constituent of giant-planet atmospheres and is used as a tracer of atmospheric structure, chemistry, and dynamics in Jupiter and Saturn [23–25]. These combustion and exoplanet applications extend the demand for accurate $NH_3$ spectroscopy beyond room-temperature terrestrial conditions. For such applications, broadening coefficients must be accurate under elevated-temperature conditions, where transitions corresponding to higher rotational-energy states become increasingly significant [26].

The high-resolution transmission molecular absorption database (HITRAN) has been the primary source of molecular absorption parameters for atmospheric and planetary radiative-transfer applications. $NH_3$ is listed as molecule 11 in the HITRAN database, and the current HITRANonline line-by-line interface lists $NH_3$ with more than $10^5$ transitions spanning the far-infrared to near-infrared spectral range [19]. In HITRAN-based simulations, an isolated transition is represented by a normalized line-shape function, commonly a Voigt profile, which combines thermal Doppler broadening with collisional Lorentz broadening, determined by pressure-broadened half-widths. For each transition, HITRAN stores self- and air-broadening parameters ($\gamma_{\mathrm{self}}$ and $\gamma_{\mathrm{air}}$), making their accuracy central to simulated absorption coefficients [19].

Broadening coefficients of $NH_3$ in HITRAN historically relied on empirical rotational quantum-number-based correlations. Several such empirical descriptions of $NH_3$ pressure-broadened half-widths have been developed. For instance, Bleaney and Penrose [27] proposed a simple power-law expression, but it was unable to reproduce the measured linewidths satisfactorily and broke down for some limiting cases. Taylor [24] later attempted to represent Varanasi's self-broadened linewidths [28] using a cubic expression, but the explicit formula and fitted constants were not reported. Brown and Peterson [29] subsequently proposed an empirical expression for the pure-rotational transitions measured in their work and for the $\nu_1$- and $\nu_3$-band data of Pine et al. [30]; however, comparisons with Nemtchinov's measurements showed deviations as large as 10% at high rotational-energy states [31]. These limitations motivated Nemtchinov to introduce a fourth-degree polynomial; later, the same group reported an improved and more compact polynomial representation of the same linewidth trends [31]. This later expression was derived from Nemtchinov's experimental dataset and was described as an improvement over the earlier fourth-degree polynomial; it captured important rotational trends within the measured range, but direct extrapolation can produce unrealistically small or negative half-widths.

HITRAN2012-2024 [10,32–34] adopted the most recent Nemtchinov correlation [31], but avoided direct extrapolation by clamping or defaulting the broadening coefficients outside the validated range [1,19,35]. This treatment prevents negative widths, but it suppresses the observed rotational dependence and can introduce artificial discontinuities in line-by-line simulations. The need for reassessment is additionally reinforced by the expanded experimental literature. Since the Nemtchinov measurements [31], several Fourier-transform infrared (FTIR) spectrometers and laser-based studies have reported $NH_3$ broadening coefficients for multiple bands, branches, and perturbers [18,31,36–46]. These datasets provide broader and more comprehensive constraints on $\gamma_{\mathrm{air}}$ and $\gamma_{\mathrm{self}}$, including transitions and branches not directly represented in the original Nemtchinov fit [31].

The objective of this work is therefore to reassess the $NH_3$ self- and air-broadened Lorentz half-widths used in HITRAN and to develop revised empirical correlations suitable for database implementation. The new correlations are constrained to preserve physically expected rotational behavior and avoid both direct Nemtchinov over-extrapolation and HITRAN's clamping/default plateaus. The resulting correlations are evaluated statistically against measured linewidths and spectrally against measured $NH_3$ absorption spectra to determine whether line-parameter improvements propagate into practical simulations.

## 2. Theory

Because $NH_3$ is a nonlinear four-atom molecule, it has six vibrational modes. The $\nu_1$ and $\nu_2$ modes are symmetric modes and produce parallel-type bands, while $\nu_3$ and $\nu_4$ are asymmetric degenerate modes and produce perpendicular-type bands. For such a symmetric-top molecule, $J$ is defined as the rotational quantum number describing the total rotational angular momentum of the molecule. The quantum number $K$ is the projection of $J$ along the principal symmetry axis of $NH_3$. Finally, since the broadening treatment of $^{14}NH_3$ in HITRAN is identical to that of $^{15}NH_3$, "$NH_3$" will be used to collectively refer to the two isotopes. Table 1 summarizes the four fundamental vibrational frequencies of $NH_3$ ($\nu_3$ and $\nu_4$ are doubly degenerate) [10].

Table 1. Fundamental vibrational modes of $NH_3$: approximate band centers, band type, and vibrational description.

| **Band** | **Frequency ($cm^{-1}$)** | **Type** | **Description** |
|---|---|---|---|
| $\nu_1$ | 3350 | Paral. | Symmetric stretch |
| $\nu_2$ | 950 | Paral. | Symmetric bend |
| $\nu_3$ | 3450 | Perp. | Asymmetric stretch |
| $\nu_4$ | 1600 | Perp. | Asymmetric bend |

For an isolated transition centered at $\tilde{\nu}_0$, the absorption coefficient at frequency $\tilde{\nu}$ can be written as

$$\alpha(\tilde{\nu}) = NS(T)\phi(\tilde{\nu}, T, P), \tag{1}$$

where $N$ is the absorber number density, $P$ is the pressure, $S(T)$ is the line strength at temperature $T$, and $\phi(\tilde{\nu}, T, P)$ is the normalized line-shape function satisfying

$$\int_{-\infty}^{+\infty} \phi\,(\tilde{\nu}; T, P)\, d\tilde{\nu} = 1. \tag{2}$$

The Voigt line shape is usually assumed when working with the HITRAN database, which is the convolution of the Doppler-broadened Gaussian profile and the collision-broadened Lorentzian profile:

$$\phi_V(\tilde{\nu}) = \int_{-\infty}^{+\infty} \phi_G\,(\tilde{\nu} - \tilde{\nu}')\phi_L(\tilde{\nu}')\, d\tilde{\nu}'. \tag{3}$$

The Gaussian component describes thermal Doppler broadening, whereas the Lorentzian component describes pressure broadening. The Lorentzian profile is written in terms of the pressure-shifted line center $\tilde{\nu}_c$ and Lorentz half-width at half maximum $\Gamma_L$ as:

$$\phi_L(\tilde{\nu}) = \frac{1}{\pi} \frac{\Gamma_L}{(\tilde{\nu} - \tilde{\nu}_c)^2 + \Gamma_L^2}. \tag{4}$$

The Lorentz full width at half maximum is therefore:

$$\Delta\tilde{\nu}_C = 2\Gamma_L. \tag{5}$$

The Doppler width arises from the thermal velocity distribution of the absorbing molecules. The Doppler full width at half maximum is:

$$\Delta\tilde{\nu}_D = \frac{\tilde{\nu}_0}{c}\sqrt{\frac{8k_B T \ln 2}{m}}, \tag{6}$$

where $m$ is the mass of the absorber, $k_B$ is the Boltzmann constant, and $c$ is the speed of light. When the molecular weight $M$ is expressed in gmol$^{-1}$, the same expression is commonly written as

$$\Delta\tilde{\nu}_D = 7.1623 \times 10^{-7}\tilde{\nu}_0\sqrt{\frac{T}{M}}, \tag{7}$$

with $\Delta\tilde{\nu}_D$ and $\tilde{\nu}_0$ in cm$^{-1}$.

The Lorentzian component is controlled by the collision-broadened half-width. For a mixture of $NH_3$ with perturbing species $j$, the effective Lorentz half-width is:

$$\Gamma_L(T,P) = P\sum_j X_j\,\gamma_{NH_3-j}(T), \tag{8}$$

where $X_j$ is the mole fraction of perturber $j$, and $\gamma_{\mathrm{NH_3}-j}$ is the pressure-broadening coefficient in cm$^{-1}$atm$^{-1}$. In HITRAN notation, the principal broadening coefficients are $\gamma_{\mathrm{air}}$ and $\gamma_{\mathrm{self}}$, both defined as Lorentz half-widths at $T_0 = 296$ K and $P = 1$ atm.

For an $NH_3$–air mixture, the effective Lorentz half-width used in the Voigt calculation is

$$\Gamma_L^{\mathrm{NH3-air}}(T,P) = P\left[X_{\mathrm{NH_3}}\gamma_{\mathrm{self}}(T) + \left(1 - X_{\mathrm{NH_3}}\right)\gamma_{\mathrm{air}}(T)\right], \tag{9}$$

where $X_{\mathrm{NH_3}}$ is the $NH_3$ mole fraction.

## 3. Dataset compilation

The present dataset was assembled from published laboratory measurements of $NH_3$ pressure-broadened half-widths. The compiled dataset contains 1317 measured self-broadened and 1231 measured air-broadened half-widths at or reduced to 296 K, spanning $J''$ = 0–16. In most cases, air broadening linewidth was not measured directly, but rather inferred from measured $N_2$- and $O_2$-broadening coefficients using the dry-air approximation:

$$\gamma_{\text{air}} = 0.79\gamma_{\text{N2}} + 0.21\gamma_{\text{O2}} \tag{10}$$

Together, the surveyed studies cover transitions corresponding to several fundamental, overtone, and combination bands including: $\nu_1$, $\nu_2$, $\nu_3$, $\nu_4$, $2\nu_2$, $2\nu_3$, $2\nu_4$, $\nu_1 + \nu_3$ (combination), $\nu_2 + \nu_4$ (combination), which is important in investigating the potential effect of vibrational dependence. All data used in the present work is compiled in the Supplementary Material (SM). A summary of the studies considered is given in the following:

- Fabian et al. [47] measured $N_2$-, $O_2$-, and air-broadening coefficients of $NH_3$ in the $\nu_2$ band using FTIR spectroscopy. The study is valuable because it provides broad $\nu_2$-band coverage with Q-and R-branch transitions. However, the subsequently published corrigendum notes that the originally reported pressure-broadening coefficients required correction by a factor of 1.333 because of a pressure-unit handling error [48]; the correction is applied throughout the present work.
- Aroui et al. [37] measured $^{14}NH_3$ line intensities, pressure-broadening coefficients, and line-mixing parameters in the $\nu_4$ band near 1550 cm$^{-1}$. The measurements used a high-resolution Bruker IFS 120 HR FTIR with a globar source. The dataset contains 57 isolated P-branch lines of $NH_3$ perturbed by $O_2$ and air at 296 K, with $J'' =$2–9 and $K'' =$0–$J''$. In a separate study, the same group reported self-broadening coefficients for $NH_3$ in the $\nu_2$ and $\nu_4$ bands, together with absolute line intensities in the $\nu_2$ band [40]. The measurements were performed at room temperature using a Bruker IFS 120 HR FTIR. The authors fitted Voigt profiles to pressure-series spectra and reported 69 self-broadening coefficients for $\nu_2$-band R-branch lines, with additional $\nu_4$-band measurements. This dataset is particularly useful for self-broadening trends with $J$ and $K$, although it contributes purely self-broadening coefficients.

- Pine and Markov [43] studied self- and foreign-gas-broadened lineshapes in the $\nu_1$ fundamental band near 3 μm using high-resolution difference-frequency laser spectroscopy. The work provides valuable rotational-quantum-number coverage and detailed lineshape information, including Dicke narrowing, speed dependence, and line mixing. The measurements included self, $N_2$, $O_2$, $H_2$, Ar, and He broadening for Q- and R-branch transitions.
- Nemtchinov et al. [31] provided the foundational broadening correlation used for $NH_3$ throughout HITRAN2012-2024 [1,10,19,35]. Their work measured absolute intensities and collision-broadened half-widths in the 10 μm region, including $H_2$-, $N_2$-, and $O_2$-broadened widths at 200, 255, and 296 K, while self- and air-broadened values were reported at 296 K. The study covered transitions from P(6,5) to R(8,8), examined the dependence on $J$ and $K$, and introduced a polynomial linewidth representation in terms of $J$ and $K$.
- Koshelev et al. [18] used an external-cavity tunable diode laser (ECTDL) spectrometer to measure $N_2$- and $O_2$-pressure broadening of $^{14}NH_3$ in the near-infrared 1.5-μm $\nu_1$+$\nu_3$ combination band. The study focused on seven isolated P($J''$,0) transitions with $J''$ =2–9, and line centers between about 6422 and 6565 $cm^{-1}$. Air-broadening coefficients were not measured directly but were calculated from the $N_2$- and $O_2$-broadened values using the dry-air mixture approximation (Eq. 10).
- O'Leary et al. [36] recorded the near-infrared spectrum of $^{14}NH_3$ between 6850 and 7000 $cm^{-1}$ using diode-laser-based cavity-enhanced absorption spectroscopy (CEAS), supported by FTIR spectra between 6400 and 6900 $cm^{-1}$. Pressure-broadening parameters were measured for 15 individual lines in air, $N_2$, $O_2$, and $NH_3$.

- Guinet et al. [38] performed a high-accuracy study of 22 R-branch transitions in the $\nu_2$ fundamental band. More than 80 pure-$NH_3$ spectra were recorded at 235, 264, and 296 K using high-resolution FTIR. The work provides self-broadening coefficients, self-shifts, and temperature-dependence information for the selected R-branch lines.
- Owen et al. [39] targeted the strong $NH_3$ absorption feature near 1103.46 cm$^{-1}$, which consists of six closely spaced R(6,K) transitions. The authors used a tunable distributed-feedback quantum cascade laser (DFB-QCL) rather than FTIR to resolve the cluster. They measured line strengths and collisional broadening coefficients of $N_2$, $O_2$, $CO_2$, and $H_2O$ at room temperature using Voigt and Galatry profiles.
- Sur et al. [41] measured line intensities and temperature-dependent broadening coefficients for nine Q(J,K) transitions in the $\nu_2$ band between 961.5 and 967.5 cm$^{-1}$. The study used two DFB-QCLs with spectra fitted using Voigt profiles. Broadening partners included Ar, $N_2$, $O_2$, $CO_2$, $H_2O$, and $NH_3$, and the temperature dependence was measured from 300 to 600 K.
- Maaroufi et al. [44] measured self-broadening, self-shift, and line-mixing parameters for $NH_3$ in the $\nu_1$ and $\nu_3$ bands in the 3050–3600 cm$^{-1}$ region. The spectra were recorded at 295 K using a high-resolution Bruker IFS125HR FTIR. About 510 rovibrational lines were analyzed, covering rotational quantum numbers $1 \leq J \leq 11$. The study included the three branches of the parallel $\nu_1$ band and the perpendicular $\nu_3$ band. In a later study [42], the same group extended the $NH_3$ broadening dataset to include $N_2$- and $O_2$-broadening, pressure-shift, and Dicke-narrowing parameters in the $2\nu_4$ band in the 3100–3600 cm$^{-1}$ region. The study included 310 rovibrational lines at 295 K, with $1 \leq J \leq 11$ and $0 \leq K \leq 11$. Voigt and soft-collision Galatry profiles were used, and air-broadening coefficients were derived from the measured $N_2$- and $O_2$-broadening coefficients.

- Alturaifi and Petersen [45] extended the Q-branch $\nu_2$-band dataset by measuring 12 transitions ($8 < J'' < 13$) between 957.5 and 963.0 cm$^{-1}$ using a DFB-QCL. The study reported line strengths and broadening coefficients for collisions with $N_2$, $O_2$, Ar, He, and $NH_3$ at 296 K.
- The study by Coxon et al. [46] is the most comprehensive recent $NH_3$ dataset in the 685–1250 cm$^{-1}$ region ($0 \leq J'' < 17$). The authors used a Bruker IFS 125HR spectrometer recorded pure- and air-broadened $NH_3$ spectra at 296 K.

Table 2 summarizes the surveyed studies used in compiling the dataset.

Table 2. A summary of the studies compiled for the dataset of this work.

| **Spectral range (cm$^{-1}$)** | **Band** | **# lines** | **P / Q / R** | $N_{\gamma_{self}}$ | $N_{\gamma_{air}}$ | **$J''$ range** | **Light source** | **Ref.** |
|---|---|---|---|---|---|---|---|---|
| ~900–1180 | $\nu_2$ | 176 | 2 / 64 / 71 | 0 | 176 | 0–15 | FTIR | [47]* |
| 1474.7–1595.1 | $\nu_4$ | 57 | 57 / 0 / 0 | 0 | 57 | 2–9 | FTIR | [37] |
| 1068.0–1595.1 | $\nu_2$ and $\nu_4$ | 131 | 64 / 0 / 67 | 131 | 0 | 2–12 | FTIR | [40] |
| 851–1118 | $\nu_2$ | 91 | 50 / 0 / 41 | 91 | 91 | 0–8 | FTIR | [31] |
| ~3331–3440 | $\nu_1$ | 74 | 0 / 63 / 11 | 74 | 74 | 0–9 | Difference-frequency laser | [43] |
| 6422.4–6564.9 | $\nu_1 + \nu_3$ (combination) | 7 | 7 / 0 / 0 | 0 | 7 | 2–9 | External-cavity tunable diode laser | [18] |
| 6900–7000 | $2\nu_3$ | 2 | 0 / 0 / 2 | 0 | 2 | 3 | CEAS with external-cavity diode laser; FTIR support | [36] |
| 1126.0–1171.4 | $\nu_2$ | 22 | 0 / 0 / 22 | 22 | 0 | 9–11 | FTIR; tunable diode laser check | [38] |
| 1103.43–1103.49 | $\nu_2$ | 6 | 0 / 0 / 6 | 0 | 6 | 6 | DFB-QCL | [39] |
| 961.5–967.5 | $\nu_2$ | 9 | 0 / 9 / 0 | 9 | 9 | 3–10 | Two DFB-QCLs | [41] |
| 3050–3600 | $\nu_1$ and $\nu_3$ | 512 | 199 / 158 / 155 | 512 | 0 | 0–12 | FTIR | [44] |
| ~3100–3533 | $2\nu_4$ | 310 | 134 / 83 / 92 | 0 | 310 | 1–11 | FTIR | [42] |
| 957.5–963.0 | $\nu_2$ | 12 | 0 / 12 / 0 | 12 | 12 | 9–12 | DFB-QCL | [45] |
| 685–1250 | mainly $\nu_2$; $\nu_4$, $2\nu_2$, $\nu_2 + \nu_4$ (combination) | 1374 | 152 / 185 / 236 | 578 | 569 | 0–16 | FTIR | [46] |

* Values from [47] were divided by 1.333 per the published corrigendum [48].

## 4. The application of Nemtchinov's correlation in HITRAN

### 4.1 Nemtchinov's correlation

Nemtchinov et al. [31] reported one of the most widely used empirical representations for ammonia pressure-broadened half-widths in the 10-μm region. The measured Lorentzian half-widths were obtained from fits of Voigt profiles to high-resolution FTIR spectra. Measured half-widths of prepared $NH_3$ mixtures were represented using the binary-collision mixture relation:

$$\gamma = X_{\mathrm{NH3}}\gamma_{\mathrm{self}} + (1 - X_{\mathrm{NH3}})\gamma_i, \tag{11}$$

where $X_{\mathrm{NH3}}$ is ammonia mole fraction and $\gamma_i$ is the foreign-gas broadening coefficient. Air-broadening coefficients were calculated from the measured $N_2$- and $O_2$-broadened values using the dry-air approximation (Eq. 10).

Nemtchinov et al. [31] found that transitions in the P and R branches with comparable rotational indexing had similar half-widths, supporting the use of a compact rotational correlation rather than treating every branch independently. They, thus, used a unified rotational quantum number, $m$, as $-J$ and $J + 1$ for the P and R branches, respectively. Although the $\nu_2$ fundamental band of $NH_3$ contains Q-branch transitions, these were not included in the Nemtchinov dataset [31]. Their measurements were based on FTIR spectra and selected primarily isolated $P$- and $R$-branch lines. The Q-branch manifold is spectrally congested; later laser-based studies noted that Q(J,K) transitions are difficult to characterize with FTIR because of the close proximity of neighboring lines and associated blending [41].

To represent the observed rotational trends, Nemtchinov et al. [31] adopted an empirical polynomial form in the rotational index $m$ and the lower-state projection quantum number $K$ ($K''$):

$$\gamma^{\mathrm{NEM}}(m, K) = \beta_0 + \beta_1 m + \beta_2 K + \beta_3 m^2 + \beta_4 K^2 + \beta_5 mK. \tag{12}$$

The coefficients $\beta_i$ are tabulated separately for different broadening gases, including air and self broadening as listed in Table 3.

Table 3. $\beta_i$ coefficients in Eq. 12 for self- and air-broadening [31].

| **Broadening gas** | $\beta_0$ | $\beta_1$ | $\beta_2$ | $\beta_3$ | $\beta_4$ | $\beta_5$ |
|---|---|---|---|---|---|---|
| Self | 0.4014 | −0.03915 | 0.1261 | 0.001982 | 0.002425 | −0.01128 |
| Air | 0.1073 | −0.003811 | 0.003936 | −0.000299 | 0.0000993 | 0.000150 |

Although the Nemtchinov polynomial [31] provides a compact and useful representation over the range constrained by the original measurements ($0 \leq m \leq 9$), its behavior becomes problematic when extrapolated to higher $m$ and $K''$. Figures 1(a) and 1(b) illustrate this issue by evaluating the self- and air-broadening correlations, respectively, over $0 \leq m \leq 23$ and $0 \leq K'' \leq 22$, the span of rotational quantum numbers present in HITRAN2024. Evidently, the polynomial is not globally positive. At low $K''$, the self-broadening correlation remains positive over the plotted range, but at larger $K''$ the negative $mK''$ contribution dominates the expression at high $m$. As a result, many high-$K''$ self-broadening curves decrease rapidly and eventually become negative. Indeed, the $K''$ = 7 self-broadening curve becomes negative for all $m > 16$. This produces an unphysical divergence between low- and high-$K''$ behavior at large $m$: low-$K''$ curves remain positive or turn upward, whereas high-$K''$ curves collapse toward negative values. Additionally, the correlation does not globally ensure $\gamma_{\text{self}}$ increase with $K''$ at a given $m$, for $m > 11$. The crossing of the constant-$K''$ curves through one another and through zero is therefore not a feature of $NH_3$ collisional physics, but an artifact of extrapolating a low-to-moderate-$m$ empirical polynomial outside the region for which it was established.

The air broadening correlation shows a different but equally important instability; the curves decrease monotonically with $m$, and several $K''$ curves cross zero at moderate-to-high $m$. For example, the $K'' = 0$ air-broadening curve becomes negative for all $m > 13$. This extrapolation problem is central to the present reassessment. The Nemtchinov correlation [31] is valuable because it captured the measured rotational dependence of $NH_3$ half-widths in a compact database-ready form. However, the same polynomial form contains no positivity constraint and no physically motivated high-$m$ asymptote. Consequently, extending it to high-$J''$ transitions can generate unrealistically small or negative half-widths. This limitation explains why later HITRAN implementations did not simply apply the Nemtchinov polynomial [31] to all high-$J''$ transitions, but instead used fixed or clamped values outside the validated range.

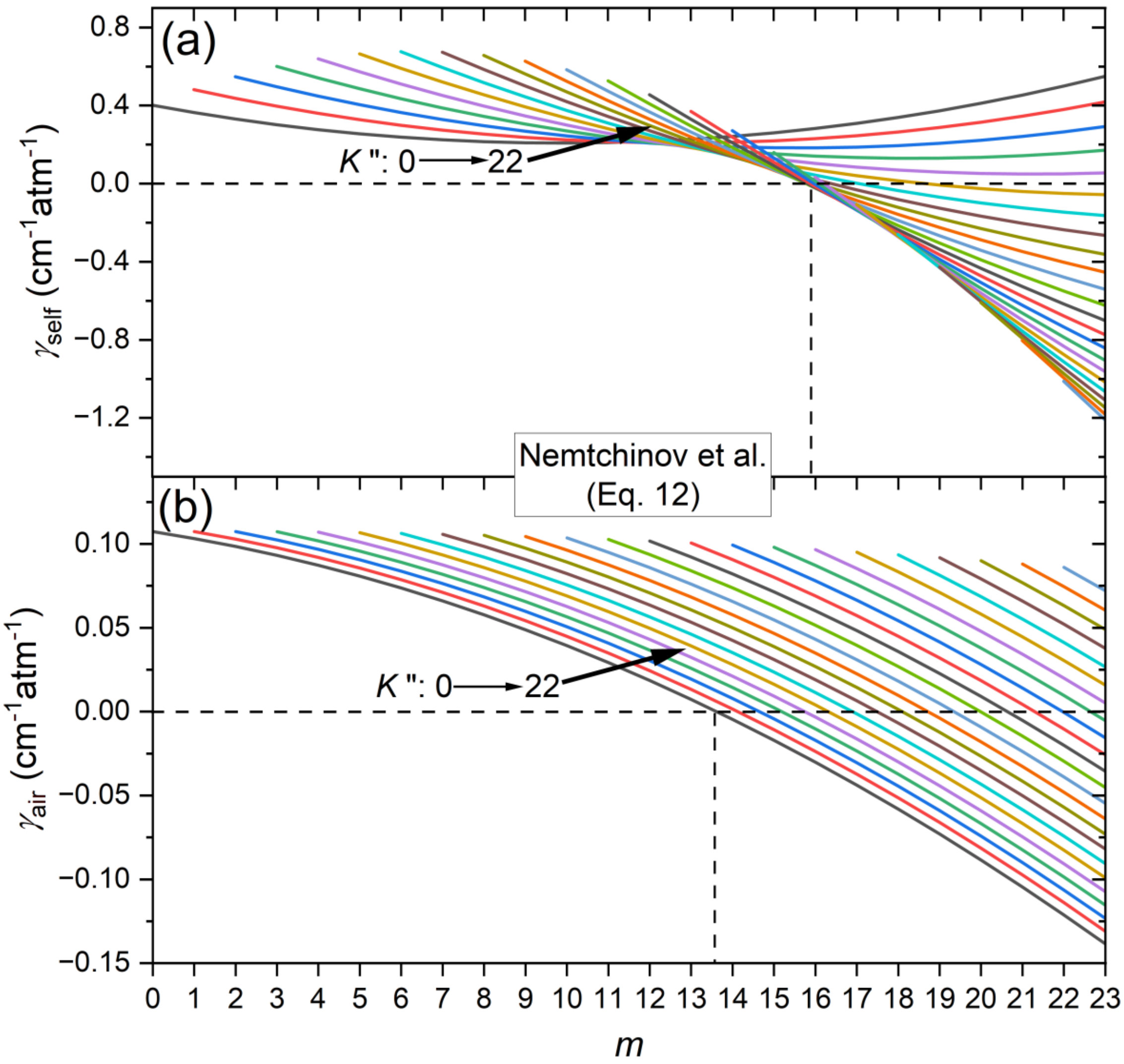

Figure 1. Evaluation of the Nemtchinov correlation [31] (Eq. 12) for (a) self- and (b) air-broadening over $0 \leq m \leq 23$ and $0 \leq K'' \leq 22$.

### 4.2 HITRAN broadening coefficients

HITRAN stores pressure-broadened Lorentz half-widths using the parameters $\gamma_{\text{self}}$ and $\gamma_{\text{air}}$, reported in units of $cm^{-1}atm^{-1}$. These parameters represent the self- and air-broadened half-widths at the reference temperature, 296 K, and are used in line-by-line simulations. In the HITRAN notation, these fields are stored as gamma_air and gamma_self. Their implementations across previous HITRAN editions are summarized in Table 4.

Table 4. A summary of $\gamma_{self}$ and $\gamma_{air}$ treatment in different editions of HITRAN.

| Edition | $\gamma_{\text{self}}$ | $\gamma_{\text{air}}$ |
|---|---|---|
| Pre-HITRAN1996 [49] | 0 | 0.075 (mostly) |
| HITRAN1996 [49] | The Brown–Peterson [29] polynomial was applied to all ammonia transitions with $J'' < 13$: $J'' < 13: \gamma = a_0 + a_1J'' + a_2K'' + a_3(J'')^2 + a_4J''K''$ For $J'' \geq 13$: $\gamma_{\text{self}} = 0.40$ | $\gamma = a_0 + a_1J'' + a_2K'' + a_3(J'')^2 + a_4J''K''$ [29] For $J'' \geq 13$: $\gamma_{\text{air}} = 0.06$ |
| HITRAN2000 [50] | $\gamma = a_0 + a_1J'' + a_2K'' + a_3(J'')^2 + a_4J''K''$ [29] | Foreign-gas widths were based on expressions derived from Nemtchinov's measurements and were applied to all $NH_3$ transitions [31]: $\gamma_{\text{foreign}} = b_0 + b_1m(m+1) + b_2k^2 + b_3[m(m+1)]^2 + b_4k^2m(m+1) + b_5k^4$, where $k = K''$ |
| HITRAN2004 [51] and HITRAN2008 [52] | No change | No change |
| HITRAN2012 [10] | Nemtchinov [31] (Eq. 12). If lower-state energy was not determined: $\gamma_{\text{self}} = 0.45$. | Nemtchinov [31]: (Eq. 12). If lower-state energy was not determined: $\gamma_{\text{air}} = 0.065$. |
| HITRAN2016 [1] | Retained the Nemtchinov polynomial [31] (Eq. 12), but emphasized that it was used only over the lower rotational range. For higher-$J$ transitions, limiting/default values were used rather than unrestricted extrapolation. For high-$J$ (> 8) transitions, mostly: $\gamma_{\text{self}} = 0.5$. | Retained the Nemtchinov polynomial [31] (Eq. 12) over the lower-$J$ range. For $J \geq 9$ and $K \leq 9$, coefficients were fixed at the corresponding $J = 9$ value. For $J \geq 9$ and $K > 9$: $\gamma_{\text{air}} = 0.0906$. |

| HITRAN2020 [35] and HITRAN2024 [19] | No change | No change |
|---|---|---|

The main source of the HITRAN2024 $NH_3$ broadening treatment is the Nemtchinov et al. [31] polynomial (Eq. 12). In the original Nemtchinov notation, the index, $m$, was defined as $m = -J$ for P-branch transitions and $m = J + 1$ for R-branch transitions. Gordon et al. defined the rotational index, $m$, used in the Nemtchinov correlation [31] adopted in the HITRAN database as $|-J, J, J + 1|$ for the P, Q, and R branches, respectively. Nevertheless, the broadening values for P-branch transitions listed in the HITRAN2024 database correspond to the Nemtchinov correlation [31] with $m = J$. The HITRAN use of $m = J$ for P-branch transitions results in closer predictions of measured data, in comparison with the use of $m = -J$ proposed by Nemtchinov et al [31].

The HITRAN treatment of self- and air-broadening coefficients can be summarized as a limited application of the Nemtchinov polynomial [31], followed by clamping or default assignment outside the intended range. The HITRAN documentation states that for $NH_3$, the air-broadening functions are fixed at the $J = 9$ value for transitions with $J \geq 9$ and $K \leq 9$, while a constant value of 0.0906 $cm^{-1}atm^{-1}$ is used for $J \geq 9$ and $K > 9$. For self-broadening, the commonly documented high-$m$ default is $\gamma_{\text{self}} = 0.500$ $cm^{-1}atm^{-1}$, while the corresponding high-$K''$ air default is $\gamma_{\text{air}} = 0.0906$ $cm^{-1}atm^{-1}$.

A direct audit of the HITRAN $NH_3$ line list shows that the actual implementation is more complicated than a single rule. The analysis of the HITRAN line list shows that, for $K'' \geq 10$, HITRAN uses fixed default values of 0.500 $cm^{-1}atm^{-1}$ for $\gamma_{\text{self}}$ and 0.0906 $cm^{-1}atm^{-1}$ for $\gamma_{\text{air}}$. For $K'' < 10$ and $m \leq 9$, most values are calculated from the Nemtchinov polynomial [31] evaluated at $(m, K'')$. For $K'' < 10$ and $m > 9$, air broadening is generally clamped to $\gamma_{\text{air}}^{\text{Nem}}(9, K'')$, whereas self broadening is mixed between the fixed default 0.500 $cm^{-1}atm^{-1}$ and the clamped value $\gamma_{\mathrm{self}}^{\mathrm{Nem}}(9, K'')$.

The resulting coefficients are therefore not equivalent to extrapolation of Nemtchinov's correlation [31], nor is it equivalent to a single default applied to all high-$m$ transitions. It is a hybrid database rule. This distinction matters because the different treatments create discontinuities in the $(m, K'')$ surface. A transition just inside the calculated region can be assigned a value from the polynomial, a transition at slightly higher $m$ can be assigned the $m = 9$ clamped value, and another transition with similar $m$ and $K''$ can be assigned a fixed default. These changes are not smooth functions of rotational quantum number and therefore do not necessarily preserve the measured rotational trends observed in the laboratory data.

The discrepancy is most evident for self broadening. The default value of 0.500 $cm^{-1}atm^{-1}$ is close to the Nemtchinov self value for some $m - K''$ combinations, especially near $K'' = 7$, but it is not representative of the full $K''$-dependent surface. For low $K''$, the clamped Nemtchinov [31] self values at $m = 9$ are substantially lower than 0.500 $cm^{-1}atm^{-1}$. For example, the Nemtchinov [31] self values at $m = 9$ increase from about 0.210 $cm^{-1}atm^{-1}$ at $K'' = 0$ to about 0.627 $cm^{-1}atm^{-1}$ at $K'' = 9$. Replacing this $K''$-dependent structure with a constant 0.500 $cm^{-1}atm^{-1}$ suppresses the rotational dependence. It overestimates some low-$K''$, high-$m$ self widths, while underestimating some high-$K''$ values that would be obtained from the clamped polynomial.

The air-broadening treatment is more systematic. For $K'' < 10$ and $m > 9$, the HITRAN values generally follow $\gamma_{\mathrm{air}}^{\mathrm{Nem}}(9, K'')$, so the dependence on $K''$ is retained but the dependence on $m$ is removed. For $K'' \geq 10$, the single default value 0.0906 $cm^{-1}atm^{-1}$ is used. Thus, high-$m$, low-$K''$ air widths are effectively flattened with respect to $m$, while high-$K''$ transitions are forced to a constant value regardless of rotational state. Although this strategy avoids the unphysical negative values produced by direct polynomial extrapolation, it also discards the measured monotonic behavior of broadening with rotational quantum numbers. A revised empirical representation should therefore not simply extrapolate the original Nemtchinov polynomial [31], but should retain its useful low-$m$ rotational dependence while imposing physically reasonable high-$m$ behavior.

One important discrepancy concerns the R branch. For most low- and intermediate-$m$ lines, the stored values correspond to the adopted convention $m = J'' + 1$ for R-branch transitions. However, a source-dependent R-branch subset instead uses $\gamma^{\mathrm{Nem}}(m-1, K'')(=\gamma^{\mathrm{Nem}}(J'', K''))$, even though the adopted convention gives $m = J'' + 1$ for R-branch transitions. In the audited HITRAN file, the $m-1$ behavior is associated with specific reference-code groups. For $\gamma_{\mathrm{air}}$, the $m-1$ pattern occurs for R-branch lines with the reference-code group "2022 4 2 1 0". For $\gamma_{\mathrm{self}}$, the same behavior occurs for R-branch lines with reference-code groups "2022 4 2 1 0" and "2022 6 2 1 0". For instance, the assigned HITRAN2024 values of the two R(1,0) transitions centered at 983.559/987.734 cm$^{-1}$ are $\gamma_{\mathrm{self}}$ = 0.364/0.331 and $\gamma_{\mathrm{air}}$ = 0.1032/0.0984 cm$^{-1}$atm$^{-1}$, respectively; the latter corresponds to $\gamma^{\mathrm{Nem}}(2,0)$, while the former corresponds to $\gamma^{\mathrm{Nem}}(1,0)$. The HITRAN treatment of self- and air-broadening coefficients for both $^{14}NH_3$ and $^{15}NH_3$ is summarized in Table 5.

Table 5. HITRAN2024 [34] treatment of self- and air-broadening coefficients for $^{14}NH_3$ and $^{15}NH_3$.

| **$J''$ and $K''$ range** | **$\gamma_{\mathrm{self}}^{\mathrm{HITRAN}}$ (cm$^{-1}$atm$^{-1}$)** | **$\gamma_{\mathrm{air}}^{\mathrm{HITRAN}}$ (cm$^{-1}$atm$^{-1}$)** | **Notes** |
|---|---|---|---|
| $J'' \leq 8$ (all branches) | Mostly $\gamma_{\mathrm{self}}^{\mathrm{Nem}}(m, K'')$ | Mostly $\gamma_{\mathrm{air}}^{\mathrm{Nem}}(m, K'')$ | Main calculated region. |
| $J'' \leq 8$ (discrepant $R$-branch transitions) | $\gamma_{\mathrm{self}}^{\mathrm{Nem}}(m-1, K'')$ | $\gamma_{\mathrm{air}}^{\mathrm{Nem}}(m-1, K'')$ | Same $m$, $K''$ in the $R$-branch can sometimes use $m-1$ ($J''$) instead of $m$ ($J''$+1). |

| $K'' < 10, J'' \geq 9$ | Mixed: either 0.500 or $\gamma_{\mathrm{self}}^{\mathrm{Nem}}(9, K'')$ | Clamped: $\gamma_{\mathrm{air}}^{\mathrm{Nem}}(9, K'')$ | Air is clamped. Self is mixed: two transitions with identical rotational quantum numbers could have different self-broadening coefficients. |
|---|---|---|---|
| $K'' \geq 10$ | 0.500 | 0.0906 | Default region. |

Finally, particular care should be taken when applying the Nemtchinov correlation [31] to Q-branch transitions. In the HITRAN implementation, Q-branch lines are treated using $m = J''$. This convention is common, but it is not directly constrained by the original Nemtchinov measurements because the Nemtchinov dataset used to construct the polynomial did not include Q-branch transitions [31].

This distinction is important because the compiled Q-branch air-broadening data indicate that the HITRAN convention does not give the best empirical mapping of the Nemtchinov polynomial to measured Q-branch widths. When the Nemtchinov air-broadening expression [31] was evaluated for Q-branch rows using $m = J''$, the comparison gave a mean absolute percentage error (MAPE) of 10.68%, and a mean signed error of +9.27%, indicating a systematic overprediction. Re-evaluating the same Q-branch transitions with $m = J'' + 1$ reduced the MAPE to 7.24%, and the mean signed error to +0.01%. Figure 2 illustrates the practical consequence of applying the Nemtchinov air-broadening correlation [31] to Q-branch transitions using the HITRAN convention ($m = J''$). The symbols show measured Q-branch $\gamma_{\mathrm{air}}$ values for two representative $K''$ values ($K''$= 2 and $K''$= 5), while the dashed curves show the HITRAN2024/Nemtchinov evaluation [31] with ($m = J''$). For both $K''$ values, the dashed curves lie systematically above the measured values, indicating that the HITRAN Q-branch implementation overpredicts the air-broadened widths. Re-evaluating the Nemtchinov expression with $m = J'' + 1$, shown by the solid curves, shifts the predicted widths downward and gives a visibly better representation. This result does not necessarily imply that the physical definition of $m$ for all Q-branch transitions should be changed. Rather, it shows that applying a P/R-derived Nemtchinov surface to Q-branch transitions using $m = J''$ introduces a systematic bias for the available Q-branch air-width measurements. Consequently, Q-branch transitions provide an additional reason to reassess the inherited HITRAN/Nemtchinov treatment and to construct a revised empirical correlation using measured Q-branch data directly, instead of relying solely on a branch convention extrapolated from the original P- and R-branch fit.

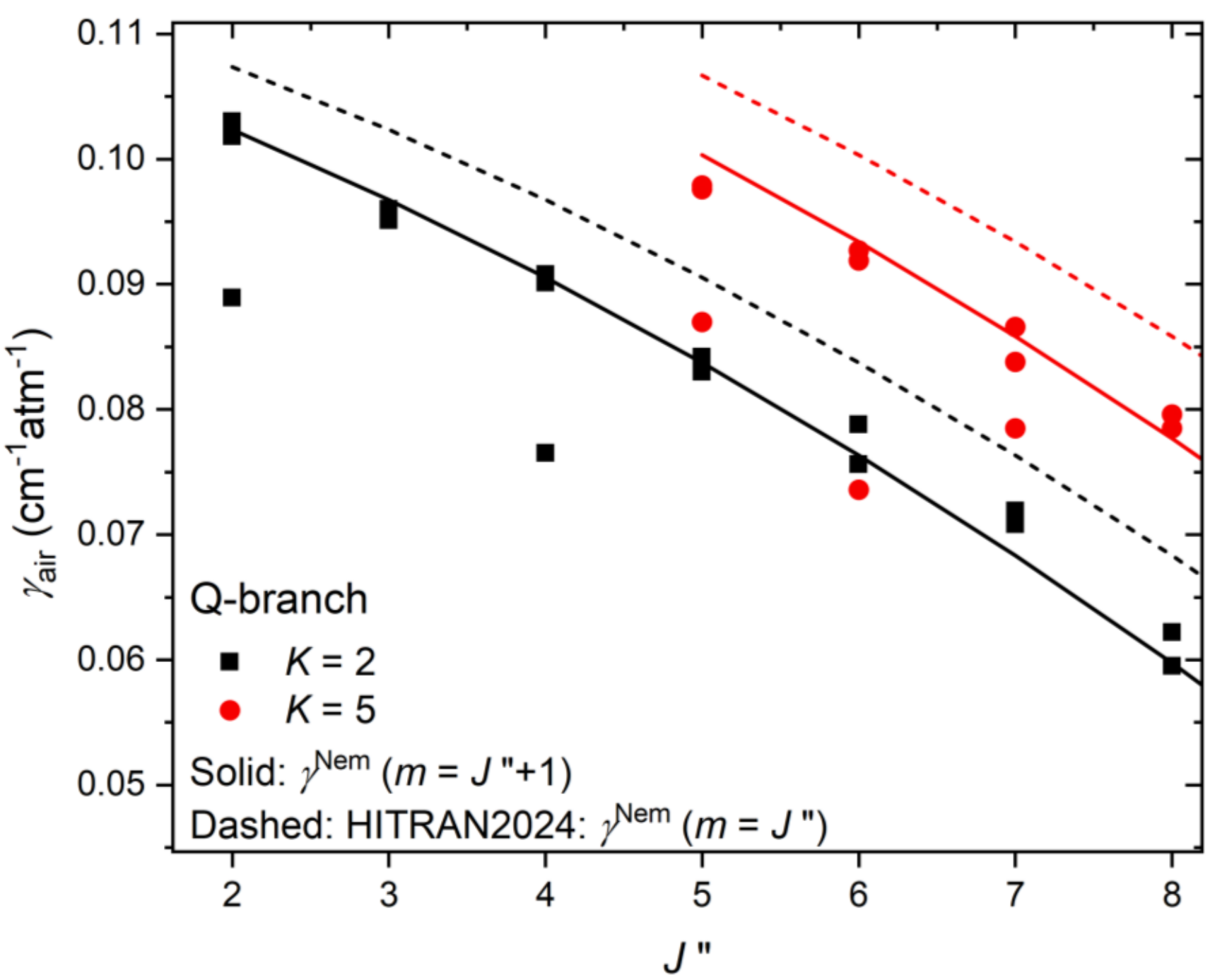

Figure 2. Comparison of measured $Q$-branch $NH_3$ air-broadened half-widths with Nemtchinov predictions [31] for $K'' = 2$ and $K'' = 5$ at different $m$. Dashed curves are the HITRAN2024/Nemtchinov implementation using $m = J''$, and solid curves show the same Nemtchinov correlation [31] evaluated with $m = J'' + 1$.

## 5. An updated $NH_3$ self- and air-broadening correlation

The Nemtchinov correlation [31] was established over two decades ago. Since its development, numerous experimental studies have measured $NH_3$ self- and air-broadened linewidths across spectral regions that include relatively high-$J$ transitions. However, no considerable updates to the ammonia self- and air-broadening coefficients have been incorporated into HITRAN since HITRAN2012 which adopted the Nemtchinov correlation [31]. This motivated the present study to develop a new correlation based on the recent body of research on ammonia pressure broadening.

**5.1 Imposed constraints**

The present empirical correlation for ammonia air- and self-broadened Lorentz half-widths was obtained by fitting the compiled laboratory measurements summarized in Table 2, given in detail in the SM. The dependent variable was the measured broadening coefficient, $\gamma^{\text{exp.}}$, and the independent variables were the branch-dependent rotational index, $m$, and the lower-state projection quantum number, $K''$. Other parameters (e.g., vibration inversion) were tested and showed no broadening dependence within the measurements uncertainty. For example, band dependence was not included as an explicit fitting variable in the present correlation because the available $NH_3$ linewidth measurements are not sufficiently uniform across vibrational bands (typically investigated across different studies) to support an independently constrained band-dependent model. This is also consistent with the HITRAN treatment of $NH_3$, which stated that the vibrational dependence of $NH_3$ pressure-broadened linewidths is expected to be smaller than the uncertainty of the measurements [10]. Accordingly, the present work follows a database-oriented approach in which $\gamma_{\text{self}}$ and $\gamma_{\text{air}}$ are represented primarily as functions of the rotational variables $m$ and $K''$, while all available measurements from different bands are used to constrain a single, physically bounded rotational surface for each broadening type.

For both self and air broadening and similar to current HITRAN adoption, the branch convention was $m = J'' + 1$ for R-branch and $m = J''$ for P- and Q-branch transitions, and in all cases, $K = K''$.

In order to ensure wide applicability of the present correlation, some constraints had to be imposed on the regression process. These were applied over the HITRAN database-use domain,

$$0 \le K'' \le J'' \le 22. \tag{13}$$

The fitted surface was required to satisfy three physical constraints over this domain. First, all predicted broadening coefficients were required to remain positive:

$$\gamma(m, K) > 0. \tag{14}$$

Second, the broadening coefficient was constrained to be decreasing with increasing $J''$ at fixed $K''$. Third, it was constrained to be increasing with increasing $K''$ at fixed $J''$. These constraints were imposed to prevent the nonphysical behavior observed when lower-order empirical forms or the original Nemtchinov polynomial [31] are extrapolated to high $J''$, including negative broadening coefficients and an unphysical steepening of the high-$J''$ self-broadening curves.

## 5.2 Uncertainty-weighted regression

In the regression process, each measurement was weighted according to its relative uncertainty. For measurement $i$, the relative uncertainty weight was

$$w_i = \left(\frac{\gamma_{\mathrm{meas},i}}{u_i}\right)^2, \tag{15}$$

where $u_i$ is the reported uncertainty assigned to the measured broadening coefficient. This is equivalent to inverse-variance weighting in relative uncertainty, because

$$w_i = \frac{1}{\left(u_i/\gamma_{\mathrm{meas},i}\right)^2}. \tag{16}$$

Thus, measurements with smaller fractional uncertainty were given larger influence in the weighted objective function, while measurements with larger fractional uncertainty were retained but contributed less strongly to the weighted performance metric.

The final model was a constrained third-degree polynomial in $m$ and $K$ ($K''$):

$$\gamma(m, K) = c_{00} + c_{10}m + c_{01}K + c_{20}m^2 + c_{11}mK + c_{02}K^2 + c_{30}m^3 + c_{21}m^2K + c_{12}mK^2 + c_{03}K^3. \tag{17}$$

Similar to Nemtchinov's correlation [31], this polynomial was used for both $\gamma_{\mathrm{self}}$ and $\gamma_{\mathrm{air}}$, with separate coefficient sets listed in Table 6. The predictions of this polynomial for $\gamma_{\mathrm{self}}$ and $\gamma_{\mathrm{air}}$ of all $NH_3$ HITRAN line list is given in the SM.

Table 6. $c_{ij}$ coefficients in Eq. 17 for self- and air-broadening.

| **Broadening gas** | $c_{00}$ | $c_{10}$ | $c_{01}$ | $c_{20}$ | $c_{11}$ | $c_{02}$ | $c_{30}$ | $c_{21}$ | $c_{12}$ | $c_{03}$ |
|---|---|---|---|---|---|---|---|---|---|---|
| Self | 5.4027E-01 | -7.3639E-02 | 1.2474E-01 | 3.3495E-03 | -8.4254E-03 | -2.2222E-03 | -5.0750E-05 | 1.8723E-04 | 0 | 1.6641E-04 |
| Air | 1.1104E-01 | -6.7307E-03 | 8.3145E-04 | -1.2694E-05 | 6.0735E-04 | -8.1533E-05 | 4.2312E-06 | -1.2694E-05 | 1.6468E-07 | -3.3881E-06 |

### 5.3 Assessment of the present correlation

Figure 3 shows the rotational dependence of the new 3rd-order polynomial fit over the $NH_3$ rotational quantum number range in HITRAN2024. The dashed boxes highlight the $m - K''$ range of the measurements used to obtain the fits. The fitted surfaces preserve the expected rotational behavior: at fixed $K''$, both self- and air-broadened half-widths generally decrease with increasing $m$, while at fixed $m$, the widths increase with $K''$. Unlike the original Nemtchinov extrapolation, the new correlations remain positive over the plotted high-$m$, high-$K''$ domain and avoid the nonphysical negative widths that motivated the reassessment. The constrained fit also removes the artificial high-$m$ steepening, producing smoothly varying curves that remain physically reasonable across the full intended range.

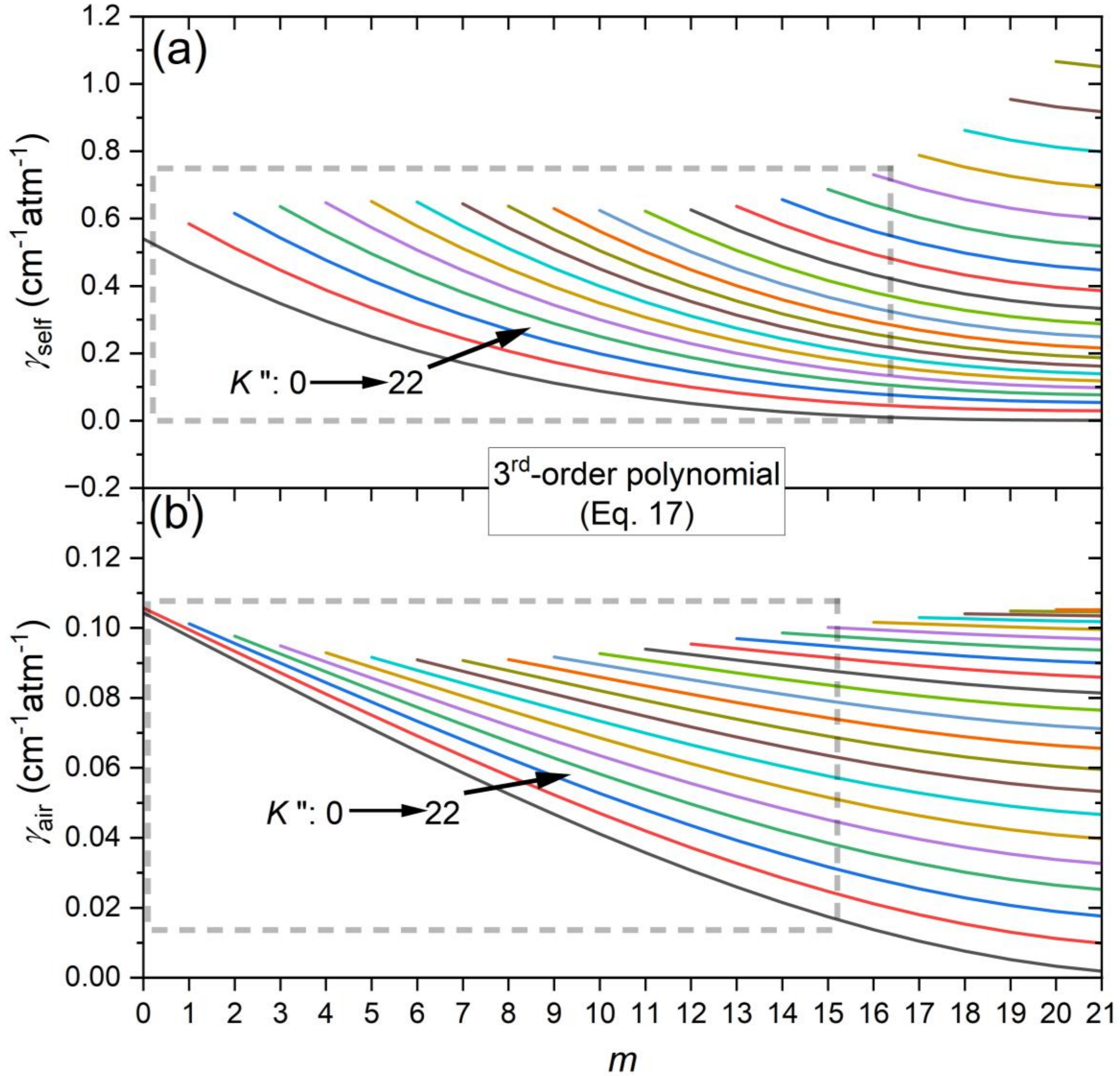

Figure 3. Evaluation of the 3rd-order correlation (Eq. 17) for (a) self- and (b) air-broadening over $0 \leq m \leq 21$ and $0 \leq K'' \leq 22$. The dashed boxes highlight the $m - K''$ range covered by the measurements used in the regression.

Figure 4(a) and 4(b) compare the constrained new-fit predictions with laser-based measurements of $NH_3$ self- and air-broadened half-widths [18,36,39,41,43,45], respectively, for $K'' = 1$–12 and $m = 1$–12. Only laser-based measurements are shown in order to provide a more internally consistent comparison and to reduce scatter associated with differences in experimental resolution and line-shape retrieval methodology. The new fit captures the overall $m$- and $K''$-dependent structure without introducing nonphysical values. The overprediction of $\gamma_{\text{self}}$ for $K'' = 10$–$12$ likely reflects the limited experimental constraint at high $K''$, rather than a failure of the overall rotational trend.

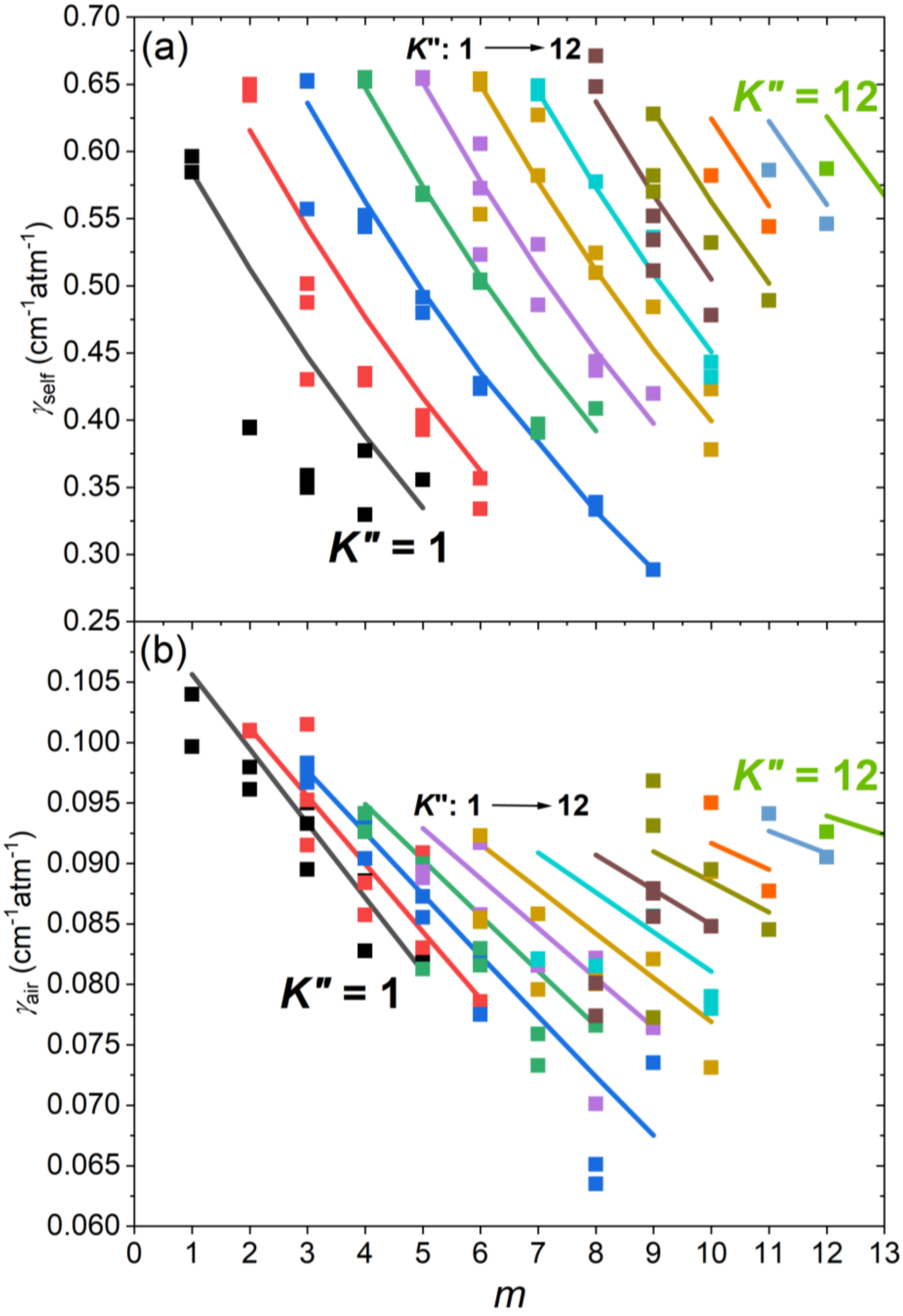

Figure 4. Comparison of laser-based $NH_3$ broadening measurements [18,36,39,41,43,45] with the new constrained third-order polynomial correlations for (a) $\gamma_{self}$ and (b) $\gamma_{air}$. Symbols represent measured half-widths, while solid lines show the new polynomial predictions for constant $K''$series from $K'' = 1$ to $K'' = 12$ over $m = 1$ – 12.

Throughout the following discussion, three schemes of self- and air-broadening coefficients will be evaluated. The first scheme is the Nemtchinov correlation [31] ($\gamma^{\text{NEM}}$, Eq. 12) extrapolated to all transitions regardless of their rotational quantum numbers (even for $J'' > 8$). The second is the HITRAN2024 listed values without change (will be denoted as $\gamma^{\text{HITRAN}}$), which is partly based on the Nemtchinov correlation [31] along with default/clamped values for high-$J$ transitions as summarized in Table 5. The third scheme is the 3rd-order polynomial correlation ($\gamma^{\text{3OP}}$, Eq. 17) introduced in this study. Two error metrics were defined to evaluate the new fit against HITRAN and the extrapolated Nemtchinov correlation as follows: for measurement $i$, the absolute percentage error was computed as

$$\epsilon_i = 100\left|\frac{\gamma_{\text{calc},i}-\gamma_{\text{meas},i}}{\gamma_{\text{meas},i}}\right|. \tag{18}$$

The first metric was the mean absolute percentage error,

$$\text{MAPE} = \frac{1}{N}\sum_{i=1}^{N}\epsilon_i, \tag{19}$$

where $N$ is the number of fitted measurements. MAPE measures the average unsigned percentage deviation between calculated and measured broadening coefficients, giving each transition equal weight.

The second was the relative-uncertainty-weighted MAPE,

$$\text{wMAPE} = \frac{\sum_{i=1}^{N} w_i\epsilon_i}{\sum_{i=1}^{N} w_i}. \tag{20}$$

The wMAPE metric emphasizes measurements with smaller relative uncertainty and is therefore a more appropriate metric for assessing agreement with the most precise data in the compiled benchmark set.

The final constrained 3rd-degree fit improved both $\gamma_{\text{self}}$ and $\gamma_{\text{air}}$ relative to HITRAN and to the Nemtchinov correlation [31]. The comparison was made using 1240 air-broadening measurements and 1317 self-broadening measurements as summarized along with the results in Table 7.

Table 7. A comparison of $MAPE$ and $wMAPE$ between predicted and measured broadening coefficients using different models.

| **Broadening gas** | **Model** | ***N*** | **MAPE (%)** | **wMAPE (%)** |
|---|---|---|---|---|
| Self | HITRAN2024 [19] | 1317 | 23.609 | 5.866 |
| | Nemtchinov [31] | | 12.126 | 5.366 |
| | 3rd-order polynomial | | 10.886 | 3.775 |
| Air | HITRAN2024 [19] | 1231 | 10.949 | 8.641 |
| | Nemtchinov [31] | | 9.276 | 8.422 |
| | 3rd-order polynomial | | 6.798 | 3.626 |

For self broadening, the new fit reduced MAPE from 23.609% for HITRAN and 12.126% for Nemtchinov to 10.886%. The reduction relative to HITRAN is large, reflecting the systematic error introduced by fixed or clamped self-width values. The improvement relative to Nemtchinov is smaller in unweighted MAPE but remains important because the constrained degree-3 fit also avoids the nonphysical negative or steep high-$J''$ behavior associated with unconstrained extrapolation. In the weighted metric, the new fit reduced the relative-wMAPE from 5.866% for HITRAN and 5.366% for Nemtchinov to 3.775%. For air broadening, the new fit reduced MAPE from 10.949% for HITRAN and 9.276% for Nemtchinov to 6.798%. The improvement is more pronounced in the uncertainty-weighted metric where it decreased from 8.641% for HITRAN and 8.422% for Nemtchinov to 3.626% for the new fit.

To visually assess the aforementioned comparison, Figure 5 compares measured broadening coefficients with HITRAN2024, the extrapolated Nemtchinov correlation [31], and the new third-order polynomial fit predictions for both $\gamma_{\text{self}}$ and $\gamma_{\text{air}}$. The one-to-one line is shown as an eye guide. The HITRAN comparisons show clear evidence of database defaulting and clamping: many self-broadened values collapse near the fixed/default $\gamma_{\text{self}}$ levels, and the air-broadened values show step-function-like behavior rather than a continuous response to the measured widths. The Nemtchinov correlation [31] gives a more continuous prediction surface, but it still shows systematic deviations from the one-to-one line, especially for self broadening and at the upper end of the air-broadening range. It also demonstrates significant underprediction of a multitude of self-broadening values, likely due to the limited number of high-$J$ transitions used in obtaining the Nemtchinov correlation [31]. In contrast, the new fit predictions are more tightly clustered around the one-to-one line for both broadening types, indicating that the constrained third-order polynomial better reproduces the measured rotational dependence while avoiding the artificial plateaus present in HITRAN and the extrapolation limitations of the original Nemtchinov expression [31].

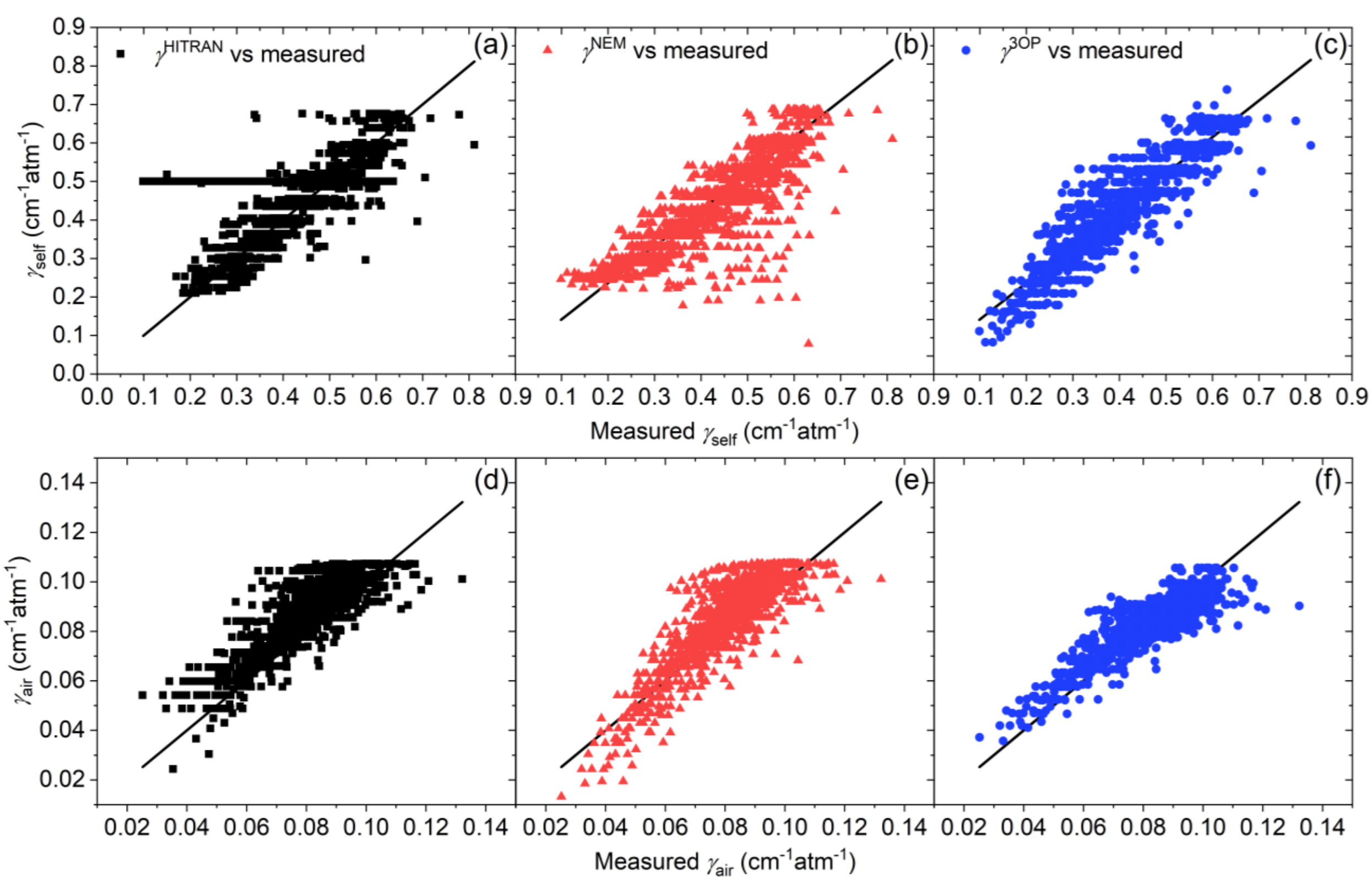

Figure 5. Comparison of measured $NH_3$ broadening coefficients with values predicted by HITRAN2024 [19], the extrapolated Nemtchinov correlation [31], and the new third-order polynomial correlation. Panels (a–c) compare $\gamma_{self}$, while panels (d–f) compare $\gamma_{air}$. The solid diagonal line is an eye guide of the identity line.

### 5.4 High-*J* (*J* > 8) transitions

Figure 6 illustrates the distribution of the strongest 600 transitions across lower-state rotational levels at 300 and 1450 K as per HITRAN2024. Governed by the Boltzmann distribution, the strongest $NH_3$ lines at 300 K are concentrated mainly at lower and moderate $J''$, where the broadening coefficients are more likely to remain within the range calculated using the Nemtchinov correlation [31]. At 1450 K, which is relevant for both combustion applications and exoplanets exploration, the distribution of the strongest lines shifts substantially toward higher $J''$, meaning that a much larger fraction of the spectroscopically important absorption is controlled by transitions for which HITRAN uses clamped or default $\gamma_{\text{self}}$ and $\gamma_{\text{air}}$ values. An exoplanet example is WASP-43b, a hot Jupiter with an equilibrium temperature of ~1440 K [53] where recent analyses reported evidence of $NH_3$ in its thermal emission spectrum [54]. Therefore, errors in the

high-$J''$ broadening treatment are not only a low-intensity transitions issue; they strongly affect high-temperature $NH_3$ present in several relevant scenarios. This is especially important for combustion diagnostics, where $NH_3$ is measured at elevated temperatures in chemical reactors

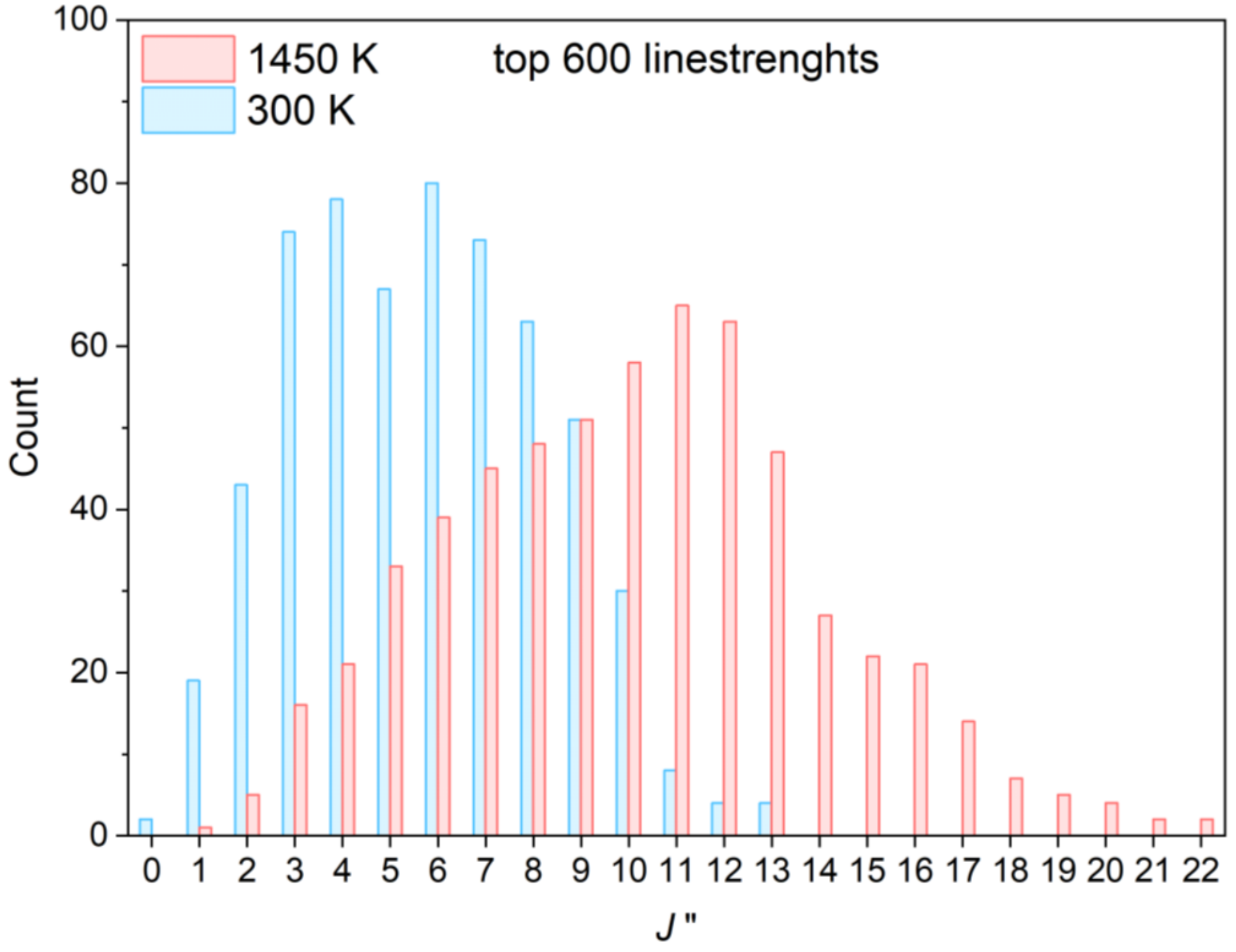

Figure 6. Distribution of lower-state rotational quantum numbers $J''$among the 600 strongest $NH_3$ transitions predicted by HITRAN2024 [19] at 300 $K$ and 1450 $K$.

[17]. As a result, high-$J''$ defaulting/clamping becomes more influential for such applications. Indeed, for transitions with $J''$ > 8, the MAPE decreases from 17.872% with HITRAN2024 to 8.305% with the new 3rd-order polynomial, while the wMAPE decreases from 12.725% to 4.400%. The improvement is even larger for $\gamma_{\text{self}}$. For $J'' > 8$, HITRAN2024 gives a MAPE of 55.295%, whereas the third-order polynomial reduces it to 10.620%. The wMAPE similarly decreases from 14.287% to 2.580%.

Figure 7 compares measured $\gamma_{\text{air}}$ with corresponding $\gamma_{\text{air}}^{\text{HITRAN}}$, $\gamma_{\text{air}}^{\text{NEM}}$, and $\gamma_{\text{air}}^{\text{3OP}}$ predictions for $K''$ = 6 R-branch transitions. The Nemtchinov correlation [31] follows the measurements reasonably near the lower-$J''$ end, but then decreases too steeply and eventually approaches negative values when extrapolated to higher $J''$. HITRAN2024 avoids this negative extrapolation by clamping the coefficient at the $m = 9$ value, but this produces an artificial plateau that removes the observed $J''$-dependence entirely. The new third-order polynomial fit provides an intermediate and more physical behavior: it remains positive, avoids the abrupt plateau, and preserves the gradual decline with $J''$seen in the measurements.

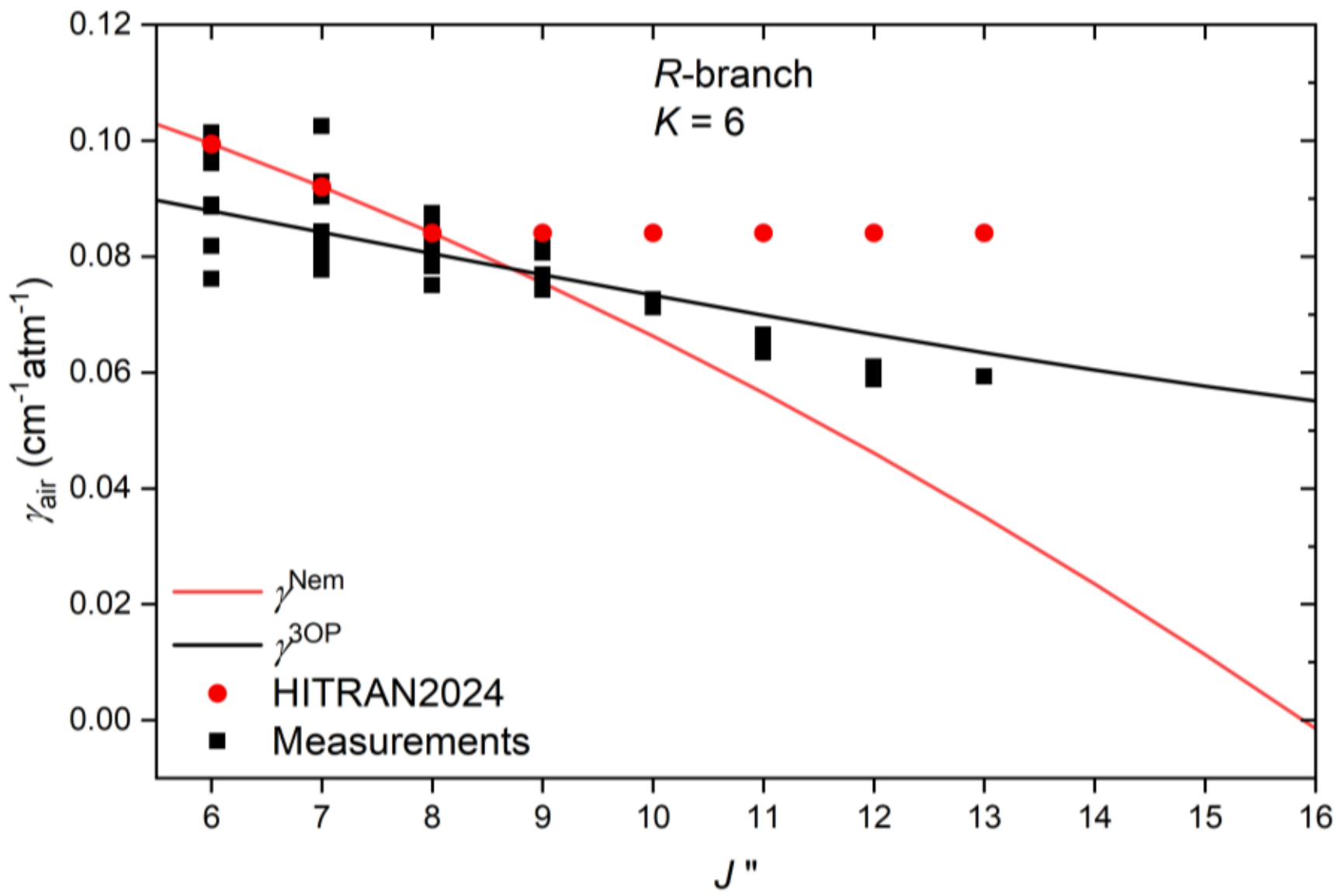

Figure 7. Comparison of measured $NH_3$ air-broadened half-widths with HITRAN2024, the extrapolated Nemtchinov correlation [31], and the new third-order polynomial correlation for $R$-branch transitions with $K'' = 6$ over $J''$ = 6 – 13.

## 6. Implications on predictions of measured $NH_3$ absorption spectra

To evaluate whether the revised broadening coefficients improve practical spectral simulations, two complementary comparisons were performed. First, the calculated spectra were compared with the spectral database of the Pacific Northwest National Laboratory (PNNL) at atmospheric pressure. Because the PNNL spectrum corresponds to dilute $NH_3$ mixtures, the Lorentz width is dominated by the foreign-gas contribution rather than by self broadening. Therefore, this comparison primarily tests the quality of the updated $\gamma_{\mathrm{air}}$-based broadening treatment after conversion to an effective $\gamma_{\mathrm{N_2}}$. The PNNL spectrum provides a band-level assessment because it tests whether improvements in individual broadening coefficients propagate to measurable improvements in simulated absorption profiles at atmospheric conditions.

A second comparison was made using the pure-$NH_3$ spectra of Alturaifi and Petersen [45]. In this case, the absorber and perturbed are both $NH_3$, so the Lorentz width is governed by $\gamma_{\mathrm{self}}$. This comparison therefore isolates the self-broadening component and provides an independent test of whether the revised $\gamma_{\mathrm{self}}$ correlation improves pressure-broadened line shapes. Together, the dilute $NH_3$ PNNL spectra and the pure-$NH_3$ spectra test the two central outputs of the present work: the revised self- and air-broadening behavior.

Finally, line strengths ($S$) of $NH_3$ transitions in the HITRAN database carry uncertainties mostly between 1-20% (HITRAN relative error codes 3-7), which is consistent with experimental observations [45]. Therefore, instead of using line-strength values stored in HITRAN, measured $S$ values [41,43,45,46] have been used in the spectral simulations in order to isolate the effects of calculated self- and air-broadening coefficients ($\gamma^{\mathrm{HITRAN}}$ and $\gamma^{\mathrm{3OP}}$) on the predictions of measured $NH_3$ PNNL spectra. Table 8 summarizes spectroscopic parameters of four spectral segments used for the comparison of PNNL predictions along with errors between predicted and experimentally measured broadening coefficients. In Table 8, $\Delta = \frac{\gamma^{\mathrm{pred.}} - \gamma^{\mathrm{exp.}}}{\gamma^{\mathrm{exp.}}}$ where the superscript and subscript of $\Delta$ correspond, respectively, to the used broadening scheme (pred. = 3OP or HITRAN) and self- or air- broadening coefficient.

Table 8. Spectroscopic parameters and broadening-coefficient errors for the $NH_3$ transitions used in the spectrum-level comparison. The relative errors correspond to self- and air-broadened half-widths obtained using either the third-order polynomial (3OP) correlation or the HITRAN2024 values.

| **Segment** | **Line center (cm$^{-1}$)** | **Measured $S$ @296 K (cm$^{-1}$/(molecule·cm$^{-2}$)** | **Transition** | $\Delta_{\mathrm{self}}^{\mathrm{3OP}}$% [a] | $\Delta_{\mathrm{self}}^{\mathrm{HITRAN}}$% [b] | $\Delta_{\mathrm{air}}^{\mathrm{3OP}}$% [c] | $\Delta_{\mathrm{air}}^{\mathrm{HITRAN}}$% [d] | **Ref. ($S$, $\gamma_{\mathrm{self}}^{\mathrm{exp.}}$, $\gamma_{\mathrm{air}}^{\mathrm{exp.}}$, $\gamma_{\mathrm{O2}}^{\mathrm{exp.}}$)** |
|---|---|---|---|---|---|---|---|---|
| (a) | 957.839001 | 2.59E-20 | Q(12,12) | 6.64 | -14.82 | 1.45 | -2.16 | [45] |
| | 958.088818 | 7.02E-21 | Q(12,11) | 2.65 | -8.42 | 0.36 | 0.11 | |
| | 959.407115 | 2.46E-20 | Q(11,11) | 6.23 | -14.68 | -1.51 | -3.72 | |
| | 959.652387 | 1.37E-20 | Q(11,10) | 2.83 | -8.09 | 2.05 | 3.31 | |
| | 960.01991 | 1.54E-20 | Q(11,9) | 2.61 | 2.25 | 1.72 | 23.55 | |
| | 960.852164 | 4.37E-20 | Q(10,10) | 7.29 | -14.09 | -3.50 | -4.63 | |
| | 961.085986 | 4.94E-20 | Q(10,9) | 5.70 | -6.02 | -1.15 | 16.65 | |
| | 961.411031 | 1.41E-20 | Q(10,8) | 5.60 | 4.60 | 0.15 | 14.86 | |
| | 961.776163 | 8.27E-21 | Q(10,7) | 4.34 | 15.74 | 2.62 | 14.68 | |
| | 962.144369 | 1.00E-20 | Q(10,6) | 5.64 | 32.28 | 5.20 | 15.05 | |
| | 962.171443 | 1.48E-19 | Q(9,9) | 8.23 | -14.09 | -2.27 | 12.14 | |
| | 962.388367 | 4.27E-20 | Q(9,8) | 6.27 | -6.37 | -0.09 | 10.81 | |
| (b) | 965.353909 | 4.19E-19 | Q(6,6) | 17.47 | 22.24 | -0.76 | 15.07 | [41] |
| | 965.499418 | 1.25E-19 | Q(6,5) | 10.51 | 13.00 | -3.21 | 9.37 | |
| | 967.34634 | 5.35E-19 | Q(3,3) | 14.20 | 7.72 | -0.66 | 9.07 | |
| | 967.406807 | 1.06E-19 | Q(3,2) | 26.25 | 15.35 | -5.84 | 0.79 | |
| | 967.449096 | 2.70E-20 | Q(3,1) | 27.42 | 12.82 | 0.01 | 4.62 | |
| (c) | 1176.609514 | 2.33E-21 | R(11,4) | -12.02 | 92.31 | 13.02 | 35.97 | $\gamma_{O2}^{exp.}$: [47]* The rest: [46] |
| | 1177.431492 | 5.59E-21 | R(10,10) | -7.24 | -17.08 | 40.02 | 41.74 | |
| | 1177.590891 | 4.22E-21 | R(11,3) | -11.54 | 135.85 | 24.87 | 51.96 | |
| | 1178.277455 | 2.01E-21 | R(11,2) | -24.54 | 160.42 | 1.22 | 26.12 | |
| | 1178.660624 | 1.89E-21 | R(11,1) | -32.12 | 240.14 | 18.55 | 53.12 | |
| | 1178.752395 | 1.87E-21 | R(12,11) | -3.15 | -4.21 | -7.21 | -5.52 | |
| (d) | 3333.02505 | 1.74E-20 | Q(6,6) | 0.00 | 4.06 | 7.53 | 24.68 | [43] |
| | 3333.89041 | 1.15E-20 | Q(4,3) | 66.11 | 59.71 | 42.14 | 56.21 | |
| | 3334.16809 | 1.14E-20 | Q(4,4) | 93.94 | 91.45 | 20.81 | 36.20 | |
| | 3334.59825 | 2.17E-20 | Q(3,3) | -2.41 | -7.95 | 1.01 | 10.89 | |
| | 3334.90128 | 1.78E-20 | Q(6,6) | 32.22 | 37.59 | 4.96 | 21.70 | |
| | 3335.47558 | 1.07E-20 | Q(5,5) | 51.60 | 54.58 | 8.37 | 24.34 | |
| | 3335.63561 | 1.15E-20 | Q(4,3) | -14.08 | -17.39 | 3.63 | 13.89 | |
| | 3335.97599 | 1.17E-20 | Q(4,4) | 17.18 | 15.68 | 1.98 | 14.98 | |
| | 3336.39057 | 2.19E-20 | Q(3,3) | 26.86 | 19.66 | 2.53 | 12.57 | |

[a] $\Delta_{self}^{3OP} = \left(\gamma^{3OP} - \gamma^{exp.} / \gamma^{exp.}\right)_{self}$.

[b] $\Delta_{self}^{HITRAN} = \left(\gamma^{HITRAN} - \gamma^{exp.} / \gamma^{exp.}\right)_{self}$.

[c] $\Delta_{air}^{3OP} = \left(\gamma^{3OP} - \gamma^{exp.} / \gamma^{exp.}\right)_{air}$.

[d] $\Delta_{air}^{HITRAN} = \left(\gamma^{HITRAN} - \gamma^{exp.} / \gamma^{exp.}\right)_{air}$.

* $\gamma_{O2}^{exp.}$ values in segment C were divided by 1.333 per the published corrigendum [48].

## 6.1 Atmospheric $NH_3$-$N_2$ spectra

The spectral data of $NH_3$ reported by PNNL is a composite spectrum normalized to 1 ppm corresponding to 12 different $NH_3/N_2$ mixtures with $NH_3$ mole fractions ranging from 0.11% to 7.28%, with an average of 1.42%. The conditions of the considered spectrum are 25.04±0.02 °C and 760±5 Torr (1±0.0066 atm). Spectral simulations using $\gamma^{\mathrm{HITRAN}}$ and $\gamma^{\mathrm{3OP}}$ (for both self and air) were conducted to compare calculated $NH_3$ spectrum with the empirical PNNL data. To that end, the air-broadened half-widths were converted to an effective $N_2$-broadened half-widths before calculating the mixture linewidth. The conversion assumes the standard dry-air composition (Eq. 10), so that the $N_2$-broadened coefficient can be obtained as

$$\gamma_{\mathrm{N_2}} = \frac{\gamma_{\mathrm{air}} - 0.21\gamma_{\mathrm{O_2}}}{0.79}. \tag{21}$$

Here, $\gamma_{\mathrm{air}}$ is either the HITRAN air-broadened value ($\gamma_{\mathrm{air}}^{\mathrm{HITRAN}}$) or the value predicted by the new air-broadening correlation ($\gamma_{\mathrm{air}}^{\mathrm{3OP}}$), and $\gamma_{\mathrm{O_2}}$ is taken from literature measurements [41,43,45,47,55], as listed in Table 8. The final Lorentz half-width used in the simulation was then computed from the $NH_3$ mole fraction,

$$\gamma_{\mathrm{eff}} = X_{\mathrm{NH_3}}\gamma_{\mathrm{self}} + \left(1 - X_{\mathrm{NH_3}}\right)\gamma_{\mathrm{N2}}. \tag{22}$$

It should be noted that the PNNL spectra have an instrumental resolution of 0.112 $cm^{-1}$ and a spectral sampling interval of 0.06 $cm^{-1}$. Therefore, when the calculated spectra exceed the PNNL absorbance at narrow features, the discrepancy does not necessarily indicate an overprediction by the model; such narrow peaks may be partially unresolved or missed because of the relatively coarse PNNL resolution. In contrast, when a calculated peak is lower than the corresponding PNNL absorbance, the discrepancy is more likely to reflect an underprediction in the simulation, i.e., inaccurately low line strengths or high broadening coefficients. This was taken into account when selecting the spectral segments considered for comparison.

Figures 8(a-d) show the results of the spectral comparisons with PNNL corresponding to the spectral segments listed in Table 8. In the $\nu_2$ Q-branch region of Figure 8(a), the features labeled

Q(11,9), Q(10,9), and Q(9,9) are directly affected by the improvement in the air-broadening coefficients. For Q(11,9), the HITRAN air-width error is largely positive ($\Delta_{\text{air}}^{\text{HITRAN}}$ =23.55%), whereas the third-order polynomial gives $\Delta_{\text{air}}^{\text{3OP}}$ =1.72%. Similarly, for Q(10,9), the error changes from $\Delta_{\text{air}}^{\text{HITRAN}}$ =16.65% to $\Delta_{\text{air}}^{\text{3OP}} = -1.15\%$, and for Q(9,9), from $\Delta_{\text{air}}^{\text{HITRAN}}$ =12.14% to $\Delta_{\text{air}}^{\text{3OP}} = -2.27\%$. This reduction in the air-width error explains why the updated simulation follows the PNNL Q-branch peaks structure more closely than the original HITRAN calculation which shows broader, and thus lower peaks. The self-width errors are also generally improved or comparable for these lines, but because the PNNL mixtures are $N_2$-diluted and contain only a small $NH_3$ fraction, the foreign-gas component is the dominant contribution to the effective Lorentz width.

The same trend is observed in Figure 8(b). For the Q(6,6) transition, $\Delta_{\text{air}}^{\text{HITRAN}}$ =15.07%, while $\Delta_{\text{air}}^{\text{3OP}} = -0.76\%$. For Q(6,5), the corresponding values are 9.37% and $-3.21\%$. These smaller air-width errors are consistent with the improved peak heights and line shapes near 965.4 cm$^{-1}$. For the Q(3,1–3) cluster, the third-order fit also reduces or maintains small air-width errors: $\Delta_{\text{air}}^{\text{3OP}} = -0.66\%$, $-5.84\%$, and 0.01% for Q(3,3), Q(3,2), and Q(3,1), respectively, compared with HITRAN values of 9.07%, 0.79%, and 4.62%.

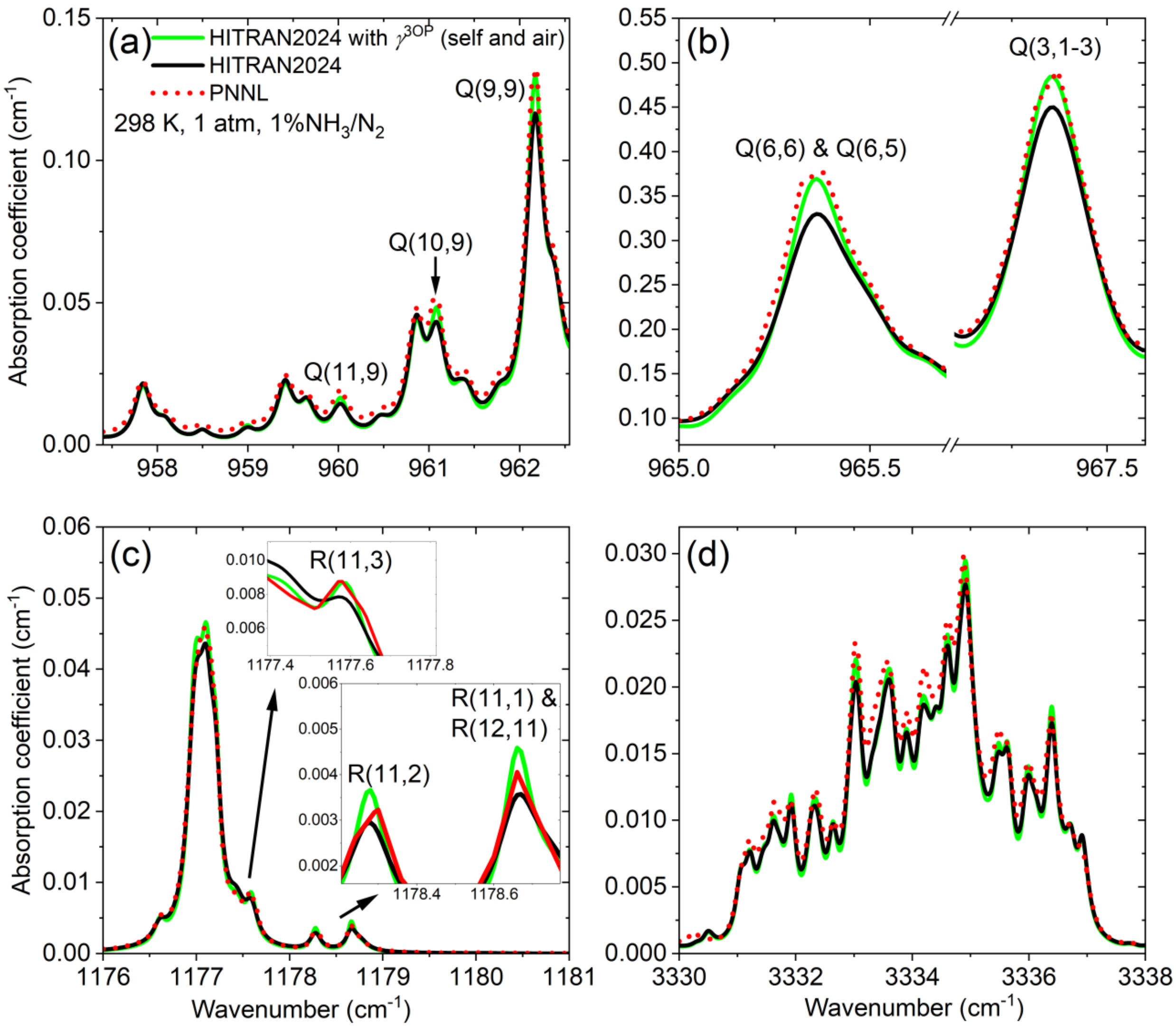

Figure 8. Comparison of measured PNNL $NH_3/N_2$ spectra [56] with simulations using the original HITRAN2024 [19] broadening coefficients and HITRAN2024 updated with the new third-order polynomial broadening correlations. (a-b) selected $\nu_2$ Q-branch transitions; (c) R-branch transitions near 1176–1181 cm$^{-1}$; (d) a congested $\nu_1$-band region near 3330–3338 cm$^{-1}$. Line strengths used in the simulations are from Refs. [41,43,45,46]. Conditions: 298 K and 1 atm.

The high-$J''$ R-branch region in Figure 8(c) provides an even clearer example of the effect of replacing fixed or extrapolated HITRAN widths in the high-$J''$ regime. For R(11,3), the HITRAN self-width error is extremely large, $\Delta_{\text{self}}^{\text{HITRAN}}$ =135.85%, while the third-order polynomial gives $\Delta_{\text{self}}^{\text{3OP}} = -11.54\%$. The air-width error is also reduced from $\Delta_{\text{air}}^{\text{HITRAN}}$ =51.96% to $\Delta_{\text{air}}^{\text{3OP}}$ =24.87%. For R(11,2), the self-width error decreases from 160.42% to $-24.54\%$, and the air-width error decreases from 26.12% to 1.22%. For R(11,1), the self-width error decreases from 240.14% to - 32.12%, and the air-width error decreases from 53.12% to 18.55%. These changes explain why the updated calculation better reproduces the weak high-$J''$ R-branch features in the inset, although residual differences remain where the fitted air-width error is still positive.

In the congested $\nu_1$-band window of Figure 8(d), the spectrum is controlled by the combined effect of many overlapping transitions; in this region. Table 8 shows only transitions with line strengths > 1E-20 $cm^{-1}/(molecule\cdot cm^{-2})$, whereas the entire line list used in the simulations is listed in the SM. The table shows that several important local contributors have substantially smaller air-width errors with the third-order fit. For example, the line at 3333.02505 $cm^{-1}$, assigned as Q(6,6), has $\Delta_{\text{air}}^{\text{HITRAN}}$ =24.68% but $\Delta_{\text{air}}^{\text{3OP}}$ =7.53%. The line at 3334.59825 $cm^{-1}$, assigned as Q(3,3), improves from $\Delta_{\text{air}}^{\text{HITRAN}}$ =10.89% to $\Delta_{\text{air}}^{\text{3OP}}$ =1.01%, and the line at 3336.39057 $cm^{-1}$, also assigned as Q(3,3), improves from 12.57% to 2.53%. These smaller air-width errors are consistent with the improved local peak and wing behavior of the updated simulation in the congested band. At the same time, some transitions in this region retain sizeable self-width deviations, so the remaining differences between the simulated and measured spectra are expected.

### 6.2 Rotationally resolved pure $NH_3$ spectra

Figures 9 (a-d) test the new self-broadening correlation using pure-$NH_3$ spectra from Alturaifi and Petersen [45]. In this case, the relevant quantity is the self-broadening error, $\Delta_{\text{self}}$, because the spectra were measured in 100% $NH_3$. In Figures 9(a) and 9(b), the improvement for the Q(12,12) and Q(12,11) lines is consistent with the table values: for Q(12,12), HITRAN underpredicts the measured self width with $\Delta_{\text{self}}^{\text{HITRAN}} = -14.82\%$, whereas the third-order polynomial gives $\Delta_{\text{self}}^{\text{3OP}}$ =6.64%; for Q(12,11), the error changes from $\Delta_{\text{self}}^{\text{HITRAN}} = -8.42\%$ to $\Delta_{\text{self}}^{\text{3OP}}$ =2.65%. This reduction in self-width error explains why the updated spectrum gives a better pressure-broadened profile for the Q(12,K) features, especially at the higher-pressure condition where $\gamma_{\mathrm{self}}$ dominates the peak width and wings.

For the Q(10,6) region in Figures 9(c) and 9(d), the table shows a different behavior: HITRAN overpredicts the measured self width by $\Delta_{\mathrm{self}}^{\mathrm{HITRAN}}$ =32.28%, while the third-order polynomial reduces this to $\Delta_{\mathrm{self}}^{\mathrm{3OP}}$ =5.64%. For Q(9,9), the HITRAN self-width error is $\Delta_{\mathrm{self}}^{\mathrm{HITRAN}} = -14.09\%$, while the third-order polynomial gives $\Delta_{\mathrm{self}}^{\mathrm{3OP}}$ =8.23%, so the new fit changes the sign of the error but keeps its magnitude moderate. This is reflected in Figures 9(c) and 9(d) where the measured peak lies above the $\gamma_{\mathrm{air}}^{\mathrm{3OP}}$-based and below the $\gamma_{\mathrm{air}}^{\mathrm{HITRAN}}$-based simulations. For Q(9,8), however, $\gamma_{\mathrm{air}}^{\mathrm{HITRAN}}$ gives the better visual match with smaller residuals. In this case the self-width errors are comparable in magnitude but opposite in sign: $\Delta_{\mathrm{self}}^{\mathrm{HITRAN}} = -6.37\%$, while $\Delta_{\mathrm{self}}^{\mathrm{3OP}}$ =6.27%. Finally, observing the residual subplots in Figures 9(a-d), the gull-wing-shaped residuals in the HITRAN curves are a diagnostic signature of an incorrect pressure-broadened line shape. Thus, the improvement is not only a reduction in peak-height error, but also a more accurate reproduction of the pressure-broadened profile shape across both the line center and wings.

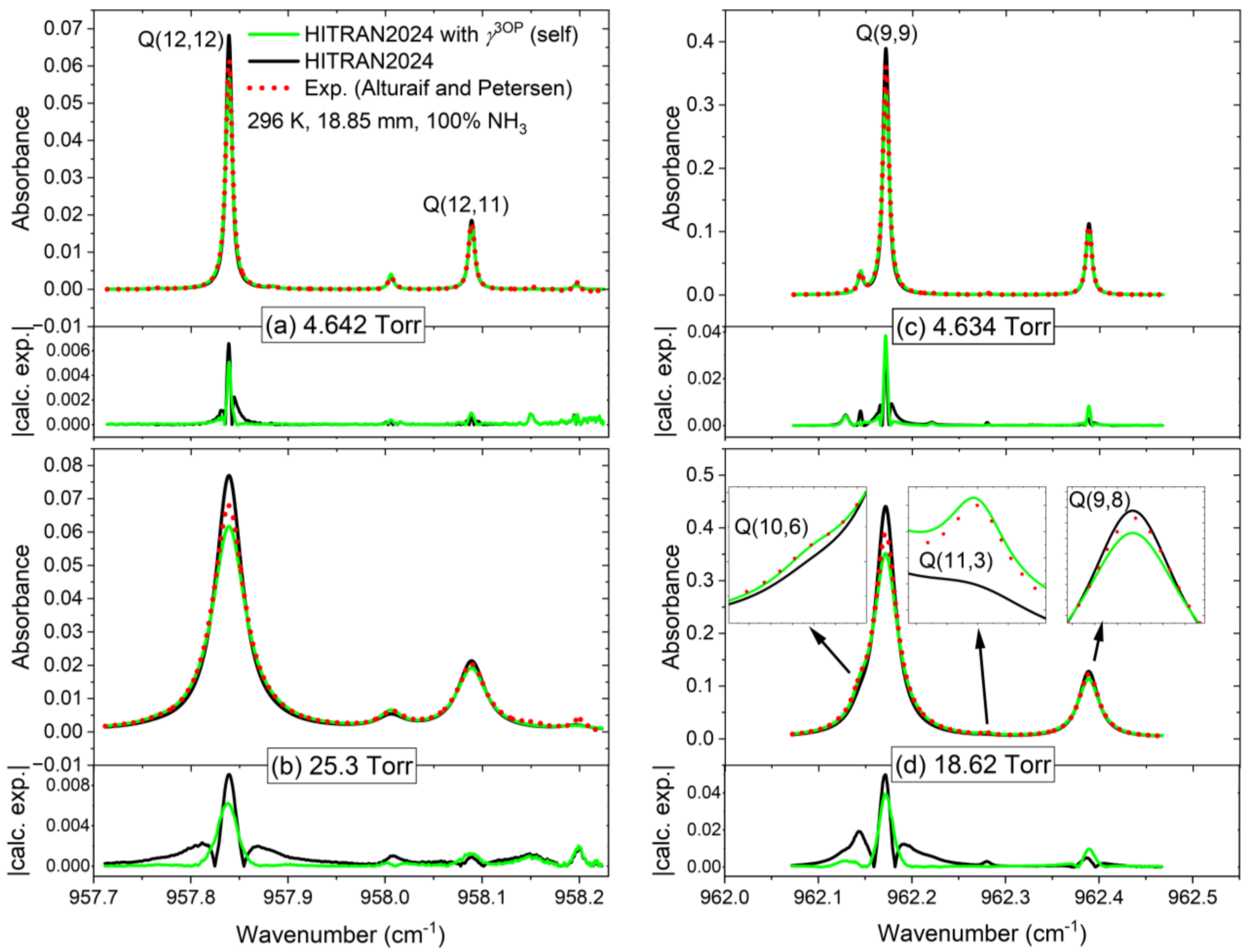

Figure 9. Comparison of pure-$NH_3$ spectra [45] with simulations using the original HITRAN2024 [19] self-broadening coefficients and HITRAN2024 updated with the new third-order polynomial correlation. The spectra were measured at 296 K, and an optical path length of 18.85 mm. (a-b) the Q(12,12) and Q(12,11) region near 958 cm$^{-1}$ at low and higher pressures, respectively. (c-d) the region near 962.3 cm$^{-1}$ at low and higher pressures, respectively. Bottom subplots show the residuals between measred and predicted spectra.

## 7. Conclusions

This work reassessed the $NH_3$ self- and air-broadened Lorentz half-widths currently used in HITRAN and developed updated empirical correlations for $\gamma_{\text{self}}$ and $\gamma_{\text{air}}$. The motivation for the reassessment is that the present HITRAN treatment is not a continuous extrapolation of the original Nemtchinov correlation [31]. Instead, it combines calculated Nemtchinov values at low and moderate rotational quantum numbers with clamped or fixed default values at higher $J''$ and $K''$. A comprehensive dataset of published $NH_3$ self- and air-broadening measurements was compiled from FTIR and laser-based studies spanning several vibrational bands and branches. The analysis showed that the dominant and most consistently supported dependence of the linewidths is rotational, rather than branch-, band-, or vibration-inversion-specific. Therefore, empirical correlations were developed for $\gamma_{\mathrm{self}}$ and $\gamma_{\mathrm{air}}$, each expressed as a constrained third-degree polynomial in the branch-dependent rotational index $m$ and the lower-state projection quantum number $K''$. The new regression procedure used relative-uncertainty weighting so that measurements with smaller fractional uncertainties contributed more strongly to the fit. The fitted surfaces were constrained to remain positive and physically reasonable over the intended HITRAN rotational domain. The new correlations improve agreement with measured linewidths relative to both HITRAN2024 and the original Nemtchinov expression [31]. For $\gamma_{\mathrm{air}}$, the new fit reduced the MAPE from 10.949% for HITRAN2024 and 9.276% for Nemtchinov to 6.798%, with an uncertainty-weighted MAPE of 3.626%. For $\gamma_{\mathrm{self}}$, the new fit reduced the MAPE from 23.609% for HITRAN2024 and 12.126% for Nemtchinov to 10.886%, with an uncertainty-weighted MAPE of 3.775%. Spectrum-level tests further support the updated coefficients. Comparisons with PNNL $NH_3/N_2$ spectra primarily assessed the foreign-gas broadening component after converting $\gamma_{\mathrm{air}}$ to an effective $\gamma_{\mathrm{N_2}}$. These comparisons showed improved peak shapes and absorption profiles when the new air-broadening correlation was used. Complementary comparisons with pure-$NH_3$ spectra isolated the self-broadening contribution and showed that replacing HITRAN $\gamma_{\mathrm{self}}$ with the new correlation reduces characteristic line-shape residuals for several Q-branch transitions. The present work, hence, provides database-ready empirical correlations for updating $NH_3$ $\gamma_{\mathrm{self}}$ and $\gamma_{\mathrm{air}}$ in HITRAN.

**Acknowledgements**

The author would like to acknowledge the support from the King Fahd University of Petroleum and Minerals (KFUPM) and the Interdisciplinary Research Center of Hydrogen Technologies and Carbon Management (IRC-HTCM) through project number INHT2612.